\documentclass[12pt]{article}
\usepackage{epsfig}

\begin{document}

\title{Zero-sound in  nuclear matter with the asymmetry parameter $-1\leq \beta\leq 1$}
\author{V.A. Sadovnikova (NRC "Kurchatov Institute", St. Petersburg, INP)
}
\date{\today}
\maketitle

\begin{abstract}
 Results for the frequencies of zero-sound excitations in the isospin asymmetric nuclear                                                                       matter  are presented for the different parameters of asymmetry $-1\leq\beta\leq 1$.
The dispersion equations are constructed within the random phase approximation. We use the effective Landau-Migdal  quasiparticle interaction.
In the paper we present  zero-sound branches of the dispersion equation solutions in the symmetric, asymmetric and neutron matter. The branches correspond to the different channels of decay of the zero-sound excitation in the nuclear matter or nuclei. In the asymmetric nuclear matter we obtain three branches  of solutions on the physical and unphysical sheets of the complex frequency plane,  $\omega_{s\tau}(k,\beta)$, $\tau=p,n,np$.  We demonstrate the change of branches with $\beta$ and conversion of the three branches  into two branches in symmetric nuclear matter ($\beta=0$) and into one branch in the neutron matter ($\beta=1$). Changing the asymmetry parameter we can connect  $\omega_{sn}(k,\beta>0)$ with $\omega_{sp}(k,\beta<0)$ at all $k$ except the special interval in which there are not solutions  at $\beta=0$.
\end{abstract}

\section{Introduction}

At the present time a  lot of attention is payed to the study of the nuclear matter properties  at the different densities, asymmetry parameters, temperature, types of interactions,  the particle composition. Our main results presented in this paper are the  branches of zero-sound excitations in the asymmetric nuclear matter. There are a lot of publications describing  the different types of the excited collective states and their decays. In paper \cite{LipPe} the three types  of excited states are obtained in the framework of local isospin density approximation approach based on the density energy functional, their contribution to the energy-weighted sum rules  in nuclei are evaluated.

In the papers  \cite{Moraw1,Moraw2} the energies and damping rate of the giant excitations and the corresponding strength functions are considered in the asymmetric nuclear matter and in nuclei at different temperature.  The zero-sound dispersion equation was constructed on the basis of the non-Markovian kinetic equation. Two (isoscalar and isovector) complex modes are obtained and a new mode predicted which exist in asymmetric nuclear matter (ANM) only.

In \cite{3} propagation of the sound modes is treated  on the basis of the kinetic theory including collisions, temperature and memory effects. In \cite{4}   results for zero sound in nuclear matter obtained in the framework of Landau-Migdal theory, are applied to the giant resonances in nuclei. To calculate the width of resonances and its temperature dependence the developed   kinetic theory was used. The role of the effective nucleon-nucleon interaction in description of the giant resonances in the hot nuclei and  dependence of the energies and widths on the temperature  are investigated in  \cite{BV}. In \cite{6}  a linear response theory starting from a relativistic kinetic equations is developed within a Quantum-Hadro-Dynamics effective field picture of the hadronic phase of nuclear matter. The dispersion relations are derived, they give the sound phase velocity and the internal structure of the normal collective modes, stable and unstable.

In our paper we investigate  zero-sound excitations  in the normal cold Fermi-liquid, consisting from the neutrons and protons at the different values of the asymmetry parameter. We study how the branches of solutions change with $\beta$.  The dispersion equation for the collective excitations in the Fermi-liquid theory \cite{9,10} is considered. In this equation we pay a special attention to the analytical structure of the polarization operators which contain the logarithmic functions. The cuts related to the logarithms are formed by the energies of free particle-hole $ph$ pairs. The stable collective solutions are placed on the right of cuts on the complex frequency plane at small wave vectors $k$. When $k$ increase, there is an overlapping of collective and $ph$ modes. At these $k$ we look for solutions under the logarithmic cuts on the nearest unphysical sheet  and obtain a complex solutions placed on the unphysical sheet. The imaginary part of solutions is interpreted as a  width of excitation appeared due to admixture of the free $ph$ pairs to the collective modes and
 the transition of the part of $ph$ pairs from the process formation of the collective mode into the state of non-interacting pairs.

 Our approach was tested in the symmetric nuclear matter (SNM) where the branches of the  nucleon zero-sound, isobar zero-sound and pion excitations were calculated \cite{7,8}.

An isospin asymmetric nuclear matter is characterized by the density of the neutrons $\rho_n$ and protons $\rho_p$. An asymmetry parameter is defined as
\begin{equation}\label{1bet}
\beta=(\rho_n-\rho_p)/(\rho_n+\rho_p).
\end{equation}
 The Fermi-momenta for the protons and neutrons are
\begin{equation}\label{1a}
p_{F\,n}= \left(\frac{3\pi^2}{2}\,(1+\beta)\,\rho_0\right)^{1/3}, \quad  p_{F\,p}= \left(\frac{3\pi^2}{2}\,(1-\beta)\,\rho_0\right)^{1/3}
,\quad p_0= \left(\frac{3\pi^2}{2}\,\rho_0\right)^{1/3},
\end{equation}
where $\rho_0=$0.17\,fm$^{-3}$.
We consider the isovector external field $V_0^\tau(\omega,k) = E_0\sum_i (\tau_z)_i\,{\rm e}^{i\vec{r_i}\vec{k}}\,{\rm e}^{-i(\omega+i\eta)t}$ which generates the set of the effective fields $V^{\tau_1\tau_2}_{eff}(\omega,k)$ in medium. They satisfy the system similar to the one obtained for the isovector dipole external field  in \cite{9}. We investigate the solutions of the dispersion equation generating by this system. The frequencies  and widths of the solutions in ANM are studied as the functions of the momentum $k$ and symmetry parameter $\beta$ at  density $\rho=\rho_0$,  and the effective Landau-Migdal  quasiparticle interaction.

The effective fields $V^{\tau_1\tau_2}_{eff}(\omega,k)$ describe the two types of excitations, being generated in the matter by the external field: noninteracting (free) particle-hole pairs ($ph$-mode) $\omega_{ph}^\tau(k)$ and the zero sound collective  mode $\omega_{s}(k)$.

In SNM at small $k$, the collective solutions $\omega_s(k)$ are real, but with $k$ increasing there is an overlapping of the collective and $ph$-modes. We obtain the complex solution in the region of overlapping using the analytical structure of polarization operator.
The imaginary part ${\rm Im}\,\omega_s(k)$ characterizes the width  of the collective mode which appeared due to mixture with the free $ph$ pairs. In SNM the isospin of these pairs is not fixed.

 In SNM, besides $\omega_s(k)$, we obtain  one more branch of the collective solutions (we denote it $\omega_{s1}(k)$). It is placed on the unphysical sheet and starts at some $k=k_c$. The feature of this branch is its decay on the $ph$ pairs of the one isospin. It may be the proton or the neutron $ph$ pairs, but not the mixture.   This branch is an important element linking the solutions in  SNM and in ANM.

In ANM  three possible mechanisms of damping of the collective excitations, three decay channels are studied. We treat  separately the overlapping of the collective mode with 1) the free neutron $ph$-pairs, 2) the free proton $ph$-pairs and 3) with the nucleon pairs whose isospin is not adjusted. Three different branches of solutions are obtained: $\omega_{sn}(k,\beta)$ in the case (1); $\omega_{sp}(k,\beta)$ when the overlapping with the proton $ph$-pairs is considered (case 2) and $\omega_{snp}(k,\beta)$ in the  case (3).  These branches are calculated on the different unphysical sheets (Sect.\,3)~\footnote{The branches $\omega_{si}(k,\beta)$,
$i=n,p,np$ depend on $k$ and $\beta$ but when it is not important  we omit  $\beta$: $\omega_{si}(k)$.}.

When we investigate the zero sound dispersion equations, the several branches of the collective solutions can be obtained. The question  appears: which of the branches can be compared with the real physical excitations, stable or unstable?  We consider as the physical sheet of the complex $\omega$-plane that one where  the real collective solutions and the energies of the free $ph$-pairs are placed on the real axis. These solutions describe the free $ph$ mode  and the stable collective modes in the matter and we name them 'the physical solutions'. The solutions at the other parameters  (the other $k$, the  densities, the asymmetry parameters, and so on) are treated as  physical ones if it is possible to do the analytical continuation over the necessary parameter  from the stable solutions to the point of the interest \cite{Newtn}. The analytical continuation  can be made to the physical or unphysical sheets. This permits us to know the nature of the every branch and to control how the branches behave  with change of $k$ and $\beta$. This is important when we are interested in the physical functions in nuclei. For example, if the strength functions of the excitations in nucleus $A$ are investigated then all physical solutions  obtained at the given $k_A$ and $\beta_A$ should be taken into account.

It is important to  note (Sect.3) that at $\beta>0$ the stable solutions belong to the branches $\omega_{sn}(k,\beta)$, at $\beta=0$ -- to $\omega_{s}(k)$ and at $\beta<0$ -- to $\omega_{sp}(k,\beta)$. It means that  at $\beta >0$ the stable solutions begin to damp with $k$ increasing due to the mixture with the neutron free pairs $\omega_{ph}^n(k)$, at $\beta <0$ due to mixture with $\omega_{ph}^p(k)$ and at $\beta=0$   the mixture with the both the proton and neutron pairs. Technically it is expressed in that at $\beta>0$ the real solutions go with $k$ increasing under the logarithmic cut of the  neutron polarization operator and we name them $\omega_{sn}(k)$.
At $\beta<0$ the real solutions go  under the logarithmic cut of proton polarization operator. This is the branch $\omega_{sp}(k)$.
At $\beta=0$ the real solutions go  under the logarithmic cuts of the both proton and neutron polarization operators. This is the branch $\omega_{s}(k)$.

In our investigation there are four special values of the wave vector: $k^p,k_t,k^{np},k_c$. At small $\beta$ we have $k^p < k_t < k^{np} < k_c$.  The character of branches, the transitions to SNM and to the neutron nuclear matter (NNM) depend of the interval where $k$ is examined. The wave vectors $k_t, k^p, k^{np}$ are functions on $\beta$  ($k_t(\beta)$ and so on) but $k_c$ is determined at $\beta=0$.

In Sect.\,2 we recall the system of equations for the effective fields which gives the dispersion equations \cite{9}.
In Sect.\,3  the discussion of the location of the solutions on the complex $\omega$-plane in SNM and ANM is presented.
 Sect.\,4 is devoted to presentation of the branches of solutions at the different $\beta$ and study the behavior  of branches at   $\beta\to 1$ and $\beta\to 0$.

\section{Dispersion equation}
We consider the isovector zero-sound excitations in the asymmetric nuclear matter.
The equations for the effective fields which are the response of ANM  to the external fields,  can be written by analogy with the paper \cite{9}.
\begin{eqnarray}
&& V^{pp}_{eff} = V_0^p + F^{pp}\,A^p\,V^{pp}_{eff} + F^{pn}\,A^n\,V^{np}_{eff} ,
\nonumber\\
\label{7}
&& V^{np} =  F^{np}\, A^p\, V^{pp}_{eff} + F^{nn}\, A^n\, V^{np}_{eff} .
\nonumber\\
&& V^{nn}_{eff} = V_0^n + F^{nn}\, A^n\, V^{nn}_{eff} + F^{np}\, A^p\, V^{pn}_{eff} ,
\nonumber\\
&& V^{pn}_{eff} =  F^{pp}\, A^p\, V^{pn}_{eff} + F^{pn}\, A^n\, V^{nn}_{eff} ,
\nonumber\\
\end{eqnarray}

$V^{pp}_{eff}$ ($V^{pn}_{eff}$) is a component of the effective field  which the proton feels when the external field $V_0^\tau$ interacts with proton  (neutron). We define $V^{np}_{eff} (V^{nn}_{eff})$  by analogy.
$A^p$, $A^n$  are the integrals over the loops of the proton and the neutron particle-hole excitations.
\begin{equation} \label{13}
A^p = A^p(\omega,k) + A^p(-\omega,k), \quad A^n = A^n(\omega,k) + A^n(-\omega,k).
\end{equation}

We use the effective Landau-Migdal interaction between the quasiparticles \cite{10}
\begin{equation}\label{17}
{\cal F}(\vec\sigma_1,\vec \tau_1;\vec\sigma_2,\vec \tau_2) = C_0\left(F + F'(\vec\tau_1\vec\tau_2) + G(\vec\sigma_1\vec\sigma_2)
 + G' (\vec\tau_1\vec\tau_2)\, (\vec\sigma_1\vec\sigma_2)\right),
\end{equation}
where $\vec\sigma$, $\vec \tau$ are the Pauli matrices in the spin and isospin spaces. $C_0= N^{-1} =\frac{\pi^2}{p_0\,m_0}$ where
$N$ is the density of states of one sort of particles, $m_0=0.94$~GeV.
In \cite{9,10} the effective interaction  between the similar  (different) particles  are $F^{nn}$, and $F^{pp}$ ($F^{np}$, $F^{pn}$). They
are associated with the  constants in (\ref{17}) by
\begin{equation}\label{3}
F^{pp}= F^{nn} = C_0\,(F + F'), \quad F^{pn} = F^{np} = C_0\,(F - F').
\end{equation}

The matrix (\ref{7}) we rewrite as
\begin{equation}\label{2}
V_{eff}= V_0 + {\cal M} V_{eff}.
\end{equation}
Here $V_{eff}$ is a column, consisting from $V^{l}_{eff}, l=pp,pn,np,nn$. The matrix ${\cal M}$ is constructed from the elements $F^{pp}\,A^p$, $F^{pn}\,A^n$ and so on in (\ref{7}). Reversing the matrix $(1-{\cal M})$ we obtain for $V_{eff}$:
\begin{equation}\label{a3}
V_{eff}= \, (1-{\cal M})^{-1}\,V_0\, =\, \frac{\tilde {\cal M}}{\det (1\, -\, {\cal M})}\,V_0,
\end{equation}
 $\tilde{\cal M}$ is the adjoint matrix.
We study the frequencies of the collective particle-hole excitations which correspond to the solutions to the dispersion equation ${\rm \det}(1\, -\, {\cal M})=0$.
\begin{equation}\label{84}
 \det(\omega,k) = (1- F^{nn}\,A^n)\,(1- F^{pp}\,A^p) - (A^p\,F^{pn})\,(A^n\,F^{np})
\end{equation}
The dispersion equation for zero-sound frequencies is defined through the determinant of the matrix coupling  the effective fields  in matter and the external field  \cite{9}:
\begin{equation}\label{8}
E(\omega,k) =  1 - C_0(F+F')A^p - C_0(F+F')A^n + 4F F' C_0^2 A^pA^n\,=\, 0.
\end{equation}
We see that  the isoscalar and isovector interactions contribute.
The similar dispersion equation was obtained in \cite{PDN} using the Bethe-Salpeter equations for RPA $ph$-propagators averaging over momentum ($A^n, A^p$).

After integration over $ph$-loops we obtain for $A^{\tau}(\omega,k)$ the expression in the form of the Migdal function \cite{9} :
\begin{equation} \label{10}
A^{\tau}(\omega,k) =\ -2\frac1{4\pi^2}\
\frac{m^3}{k^3} \left[\frac{a^2-b_\tau^2}2 \ln\left(\frac{a+b_\tau} {a-b_\tau}\right)-ab_\tau \right]\,
\end{equation}
where $a=\omega-(\frac{k^2}{2m})$, $b_\tau =\frac{kp_{F\tau}}{m}$.

\subsection {Normalization of the effective interaction Eq.(\ref{17})}
We investigate the solutions of (\ref{8}) in the asymmetric nuclear matter. A special attention is paid to the study of solutions at $\beta\to 0$ and $\beta \to 1$. When we transit  to SNM we demonstrate the  correspondence between the branches in SNM and ANM. In the neutron matter the density of the proton states goes to zero and we expect the disappearance of the branches $\omega_{sp}(k)$ and $\omega_{snp}(k)$.

To include the dependence on the density of states we change the normalization of the effective quasiparticle interactions in the particle-hole channel by the following way:
\begin{equation}\label{1c0}
C_0 \to (C_{01}C_{02})^{1/2}= (N_{1}\,N_{2})^{-1/2},\quad C_{01,02} = \frac{\pi^2}{m_0\,(p_{F1}\, p_{F2})^{1/2}}.
\end{equation}

We have included the density of states with the isospin of the $ph$ pairs before and after the interaction. Then
\begin{equation}\label{3r}
F^{pp}=  C_{0p}\,(F + F'), \quad F^{nn}=  C_{0n}\,(F + F'),
\end{equation}
$$
 F^{pn} = F^{np} = (C_{0p}\,C_{0n})^{1/2}(F - F').
$$
Here $C_{0p}= N_p^{-1}=\frac{\pi^2}{m_0\,p_{Fp}}$, $C_{0n}= N_n^{-1} = \frac{\pi^2}{m_0\,p_{Fn}}$.  We repeat the derivation of the dispersion equation (\ref{8}) with the redefined interaction (\ref{3}) and obtain the following dispersion equation
\begin{equation}\label{8r}
E(\omega,k) =  1 - C_{0p}(F+F')A^p - C_{0n}(F+F')A^n + 4F F' C_{0p}C_{0n} A^pA^n\,=\, 0.
\end{equation}

\section{Location of solutions to Eqs.\,(\ref{8}),(\ref{8r})}
In  SNM we recall the main results on the energies of the particle-hole excitations \cite{LS} and show the location of these results on the complex $\omega$-plane. Then we extend our investigation to ANM.

\subsection{SNM}
The system of equations (\ref{7}) has  two kinds of  solutions corresponding to two sorts of excitations in nuclear matter: a set of the non-interacting particle-hole pairs $\omega_{ph}(k)$ and the collective excitation $\omega_{s}(k)$.
At the beginning we consider the solution of Eq.\,(\ref{8}) in the symmetric nuclear matter. In this case $A_p(\omega,k)$ = $A_n(\omega,k)$ = $A(\omega,k)$, $A=A(\omega,k) + A(-\omega,k)$.   The dispersion equation (\ref{84}) in SNM has the form:
$$
E(\omega,k)= (1-F^{\tau\tau}A\,-\,F^{\tau\tau'}\,A)\,(1-F^{\tau\tau}A\,+\,F^{\tau\tau'}\,A)=
$$
$$
= (1-2\,C_{0}F\,A)\,(1- 2\,C_{0}F'A) = 0.
$$
It separates in two factors and there are two branches $\omega(k)$:  the branch generated by isoscalar effective interaction $F$ and the second one generated by the isovector interaction $F'$. If $F \neq F'$ the branches are separated.
Below in the text we shall study solutions of Eq.(\ref{8}) with $F=0$.

\begin{equation}\label{8s}
 1 - 2\,C_0\,F' A\,=\, 0.
\end{equation}
This equation has the solutions corresponding to the collective excitations $\omega_s(k)$.

 The excitation energies of the free noninteracting pairs are ($p<p_F$ and $|\vec p+\vec k|\geq p_F$)
\begin{equation} \label{9}
\omega_{ph}(k) = \varepsilon_{\vec p+\vec k} - \varepsilon_{p}, \quad \varepsilon_{q}= q^2/(2\,m).
\end{equation}
We  show the part of the energies of the free $ph$-pairs in Fig.\,\ref{f1}\,($left$) as the dashed area.
  In SNM we can not distinguish the excitation of the neutron particle-hole pairs and the proton ones.

The branch of the collective excitations $\omega_s(k)$ is shown by the solid curve in a schematic Fig.\,\ref{f1}. We see that the collective excitation are located above the dashed area for the small $k$. At a definite $k=k_t$ there is a overlapping of two modes. The solutions $\omega_s(k)$ become complex because of the damping of the collective mode due to admixture of $ph$-mode. In Fig.\,\ref{f1}\,($left$) at $k>k_t$ only the real part  ${\rm Re}\,\omega_s(k)$ is shown at the axis $\omega>0$  (${\rm Im}\,\omega_s(k)$ is not shown).

We can consider $\omega_s(k)$ and $\omega_{ph}(k)$ from another point of view on the complex $\omega$-plane, \\ Fig.\,\ref{f1}\,($right$).
 The  function $A$ (\ref{13},\ref{10}) has the logarithmic cuts.  At a fixed $k$ the cut of  $A(\omega,k)$ is shown in Fig.\,\ref{f1}\,($right$) by the lines (1,1').  The cut of  $A(-\omega,k)$ correspond to the line (2,2').
 \begin{equation} \label{201}
 (1,1'): -\frac{kp_F}{m} + \frac{k^2}{2m}\ \le\ \omega\ \le\ \frac{kp_F}{m}+\frac{k^2}{2m}\ ,
  \quad  (2,2'):-\frac{kp_F}{m} - \frac{k^2}{2m}\ \le\  \omega\ \le\ \frac{kp_F}{m} - \frac{k^2}{2m}\ .
\end{equation}
The point of the cut are formed by the energies of  the free $ph$-pairs $\omega_{ph}$.

The boundaries of $\omega$ in (\ref{201}) at $\omega>0$ are the boundary of the dashed area in Fig.\,\ref{f1}\,($left$). At a fixed $k$ the vertical line in the dashed area in Fig.\,\ref{f1}\,($left$) corresponds to the cut (1,1') in Fig.\,\ref{f1}\,($right$).  The every point of the solid curve in Fig.\,\ref{f1}\,($left$) at $k<k_t$ (point $A$) is  the point on the real axis on the right from the cut in Fig.\,\ref{f1}\,($right$). But the every point at $k>k_t$ in Fig.\,\ref{f1}\,($left$) (point $B$) is  the point under the cut  on the nearest lower unphysical sheet of the Riemann surface of the logarithm in  $A(\omega,k)$ (\ref{10}).
The value $k_t$ is determined by the condition that there is a solution to (\ref{8s}) such that $\omega= \omega_s(k_t) = \frac{k_tp_F}{m}+\frac{k_t^2}{2m}$, (see (\ref{201})).
Also at large $k$ we can obtain on the unphysical sheet the complex solutions with ${\rm Re}\,\omega_s(k)$  larger then the end of the cut (\ref{201}) (point $C$):   ${\rm Re}\,\omega_s(k) \ge \frac{kp_F}{m}+\frac{k^2}{2m}$.

To obtain the solutions in the overlapping region we must go under the cut to the unphysical sheet of  $A(\omega,k)$. We are failed to find solution on the physical sheet of the complex $\omega$-plane at $k>k_t$ \cite{7,8}.
The transition of solutions of (\ref{8}) to  the unphysical sheet through the cut and appearance of their imaginary part we interpret as  the damping of the collective excitations due to admixture of the free proton and neutron particle-hole pairs. In SNM we do not fix the isospin of the $ph$ pairs. The branch $\omega_s(k)$  is shown in Fig.\,\ref{f2} by the solid curves.

We have obtained the second collective branch of solutions to Eq.\,(\ref{8}) (we denote it as $\omega_{s1}(k)$). It is places on the unphysical sheet of the functions $A^p(\omega,k)$ or $A^n(\omega,k)$  but not of the both: $\omega_{s1}(k)$ is calculated on the unphysical sheet of the logarithm  of $A^n(\omega,k)$ but $A^p(\omega,k)$  is taken on the lower half plane of the physical sheet of the complex $\omega$-plane. Changing $n \leftrightarrow p$ we obtain the same $\omega_{s1}(k)$ in SNM. This branch does not appears at the real axis. It is complex, damped quickly and started at some $k=k_c$. The value of $k_c$ is obtained numerically. Branches of solutions $\omega_{s}(k)$ and $\omega_{s1}(k)$ are shown in Fig.\,\ref{f2}. At the equilibrium density $\rho=\rho_0$ we obtain $k_c=0.52\,p_0$, $k_t=0.34\,p_0$.
Here we would like to emphasize that  below it will be important that there are not solutions of type $\omega_{s1}(k)$ at $k<k_c$.

There is a question about the behavior of the solutions on the unphysical sheets when the interaction goes to zero $F'\to 0$. In this case in Landau theory the system behaves as an  ideal Fermi gas.
Our calculations demonstrate that ${\rm Re}\,\omega_{si}(k)$ and $|{\rm Im}\,\omega_{si}(k)|$ become compatible in size and quickly increase when $F'\to 0$ on unphysical sheets at every $k$. In this case we conclude that solutions disappear on the unphysical sheets.

In calculations we use $\rho_0=0.17 fm^{-3}$, $p_0=0.268\,GeV$, isovector constant $F'$ of the effective interaction Eq.\,(\ref{17}) is $F'=1.0$, effective nucleon mass $m^*=0.8\,m_0$.

\subsection{ANM}
In our model in asymmetric nuclear matter the excited  collective states  are damped due to mixture with $ph$-mode,  $\omega^\tau_{ph}$. We can distinguish the mixture with $\omega_{ph}^p(k)$, with $\omega_{ph}^n(k)$ and with $\omega_{ph}^\tau(k)$, $\tau$ is not fixed. These three variants have the separate branches of solutions.

In ANM  the dispersion equation (\ref{8r}) has the form
\begin{equation}\label{8z}
 1 - \,C_{0p}\,F' A^p\, - \,C_{0n}\,F' A^n\,=\, 0.
\end{equation}
In Eq.\,(\ref{8z}) excitations $\omega^n_{ph}(k)$ and $\omega^p_{ph}(k)$ form the cuts in functions $A^p$ and $A^n$ (Eq.\,(\ref{13}))  on the real axis.

The physical sheet and the cuts of Eq.\,(\ref{8z}) are shown in Fig.\,\ref{f3}. The cuts of $A^\tau(\omega,k)$ and $A^\tau(-\omega,k)$ ($\tau=n,p$) are:
\begin{equation} \label{12a}
 (1,1'): -\frac{kp_{F\tau}}{m} + \frac{k^2}{2m} \le \omega \le \frac{kp_{F\tau}}{m}+\frac{k^2}{2m}; \quad (2,2'): -\frac{kp_{F\tau}}{m} - \frac{k^2}{2m} \le  \omega \le \frac{kp_{F\tau}}{m} - \frac{k^2}{2m}.
\end{equation}
The cut of  $A^\tau(\omega,k)$ is labeled by $(1,1')$  and the cut of $A^\tau(-\omega,k)$ is labeled by $(2,2')$. In Fig.\,\ref{f3} two sets of cuts are shown for  $A^n$ and $A^p$  when $\beta>0$, $p_{Fn}>p_{Fp}$ and, consequently, the neutron cuts are longer than the proton ones.
In the symmetric nuclear matter the cuts are identical.

Now we consider three branches of solutions of Eq.\,(\ref{8z}) in the nuclear matter with $\beta>0$. First, we obtain the zero-sound  collective branch, it is real at small $k$. With $k$ increasing it is overlapping with the set of the neutron $ph$-pairs ($\omega^n_{ph}(k)$) and is damping due to the mixture with them. Since $p_{Fn} > p_{Fp}$ the real solutions meet the neutron cut first and go under this cut. We name this branch $\omega_{sn}(k)$.   In nuclei this damping would correspond to the semi-direct decay of the collective state due to the neutron emission.
Second, we obtain the zero-sound  branch $\omega_{sp}(k)$. It is complex  even at the small $k$ (at $\beta>0$), it is damping due to the mixture with  the set of the proton $ph$-pairs ($\omega^p_{ph}(k)$).  In nuclei this damping would correspond to the decay of the collective state due to the proton  emission. Third, there is a $\omega_{snp}(k)$, it is placed under the both proton and neutron cuts, it is damping due to overlapping with nucleon  $ph$-pairs.

\section{Solutions of the dispersion equation in ANM}
At the beginning we compare the branches of solutions in SMN and in ANM with small $\beta=0.01$.  In Fig.\,\ref{f4}$a$ we show  on the complex $\omega$-plane, the same two curves as in Fig.\,\ref{f2}  ($\omega_{s}(k), \omega_{s1}(k)$). In Fig.\,\ref{f4}$b$,$c$,$d$ we show the branches  in ANM with a small $\beta=0.01$: $\omega_{sn}(k)$, $\omega_{sp}(k)$, $\omega_{snp}(k)$. At small $\beta$ the branches in ANM are close to that in SNM, but there is not the direct correspondence between the whole branches.  It will be discussed below how the branches in ANM approximate the solutions in SNM and behave with $\beta\to0$.

In Fig.\,\ref{f4} we see that the branches start at the different $k$.
  There are four   wave vector values  which are connected with branches, Fig.\ref{f5}:\\
1) $k_t(\beta)$ - this is the value of  wave vector when the real branch of solutions becomes  complex. At $\beta=0$ the real solutions belong to $\omega_{s}(k)$, at $\beta > 0$ -- to $\omega_{sn}(k)$ and at $\beta < 0$ -- to $\omega_{sp}(k)$.\\
2) $k_c$ is the  wave vector value that the branch $\omega_{s1}(k)$ starts (Fig.\,\ref{f2}), $k_c$ and  $\omega_{s1}(k)$ exist in SNM only.\\
3) $k^{np}_{1,2}(\beta)$ are connected with $\omega_{snp}(k)$; $k^{np}_{1}(\beta=0) = k_t$. $k^{np}$ consists of two parts $k^{np}_1$ and $k^{np}_2$. Solutions $\omega_{snp}(k)$ exist at that $\beta$ and $k$ which are placed on the right and above the curve $k^{np}(\beta)$. There is not $\omega_{snp}(k)$ on the left and below of the curve $k^{np}(\beta)$. \\
4) $k^p(\beta)$ is the beginning of $\omega_{sp}(k)$  at $\beta > 0$. For the every $\beta$, $\omega_{sp}(k)$  exist at $k>k^p$.\\
We obtain that at $\beta<0.24$ there are inequalities $k^p(\beta) < k_t(\beta) < k_1^{np}(\beta) < k_c$ which are very sensitive to the input values of $F'$, $m^*$, $p_0$.

To obtain the curves at  $\beta<0$ we change the protons to neutrons (and vise verse) and reflect the curves in a vertical axis. But there is not the continuation of solutions obtained at $\beta>0$ into the ones at $\beta<0$ at all $k$.

In the Figs.\,\ref{f6}$a$ -- \ref{f6}$c$ and \ref{f7}$a$ -- \ref{f7}$c$ the branches of solutions to Eq.\,(\ref{8z}) $\omega_{sn}(k,\beta)$ (Fig.\,\ref{f6}$a$), $\omega_{sp}(k,\beta)$  (Fig.\,\ref{f6}$b$)  and  $\omega_{snp}(k,\beta)$ (Fig.\,\ref{f6}$c$) are shown for the different asymmetry parameter $\beta$. Calculations are made for the density $\rho=\rho_0$  and $\beta$= 0.01,0.2,0.5, 0.8.

\subsection{The branches $\omega_{sn}(k,\beta)$}
As it was mentioned above, at the small $k$ and $\beta > 0$ the dispersion equation has the real solutions. This takes place at $k< k_t(\beta)$ (Fig.\,\ref{f5}). We name them $\omega_{sn}(k)$ because
with $k$ increasing  there is overlapping of solutions with the neutron cut, Fig.\,\ref{f3} (point $A$). And  $\omega_{sn}(k)$ decay due to the mixture with the free neutron $ph$ pairs.

After overlapping we continue the branch of solutions  under the cut to the unphysical sheet of $A^n(\omega,k)$.
 It means that   ${\rm Im}\,\omega_{sn}(k)$ appears at the wave vectors larger the definite one ($k>k_t(\beta)$). The function  $k_t(\beta)$ is computed and shown in Fig.\,\ref{f5}.

Dependence of $\omega_{sn}(k)$ on $\beta$ is shown in Fig.\,\ref{f6}$a$.
When the asymmetry parameter grows the real part of $\omega_{sn}(k)$ increase. The dependence of $|{\rm Im}\,\omega_{sn}(k,\beta)|$ changes with $k$. We see  the special behavior of $\omega_{sn}(k)$ at small $\beta$. This is explained by the  transition to the solutions at $\beta=0$  which will be discussed below. The dotted curve show $\omega_{sn}(k)$ in NNM.

We can consider the real solutions $\omega_{sn}(k)$ at $k< k_t(\beta)$ as continuations to  $\beta>0$ of the real solutions  $\omega_{s}(k)$ existing for $\beta=0$ and  $k\leq k_t(\beta=0)$.

\subsection{The branches $\omega_{sp}(k,\beta)$}
At $\beta>0$ the branch $\omega_{sp}(k)$ is placed on the unphysical sheet completely.
Since the real solution meets the neutron cut first when $k$ increase (Fig.\,\ref{f3}) we can not say that the real solution goes under the proton cut. We construct the unphysical sheet of $A^p(\omega,k)$ similar to the sheet of $A^n(\omega,k)$ and calculate the solutions on this sheet. This sheet is the nearest unphysical logarithmic sheet of the logarithm in $A^p(\omega,k)$.
The imaginary part of $\omega_{sp}(k)$  corresponds to the damping of excitation due to mixture with the proton free $ph$-pairs,  $\omega_{sp}(k)$ is complex at the every $k$ for $\beta>0$.

In the Figs.\,\ref{f6}$b$ and \ref{f7} the branches $\omega_{sp}(k)$ are shown.
The dependence of $\omega_{sp}(k)$  on $\beta$ differs from that of $\omega_{sn}(k)$: the real part of $\omega_{sp}(k)$ decrease with $\beta$, but  $|{\rm Im}\,\omega_{sp}(k)|$ increase. In the neutron matter ($\beta=1.0$),  the  branch $\omega_{sp}(k)$ is absent, of course.
A special behavior of $\omega_{sp}(k)$ at $\beta \to 1$ is discussed in Sect.4.4.
As in Fig.\ref{f6}$a$, we see  the special behavior of $\omega_{sp}(k)$ at small $\beta$.

 Our calculations show that $\omega_{sp}(k)$ appears at the point $k=k^p(\beta)$ (Fig.\,\ref{f5}) and exists at $k\geq k^p$. This point  $\omega_{sp}(k^p)$ coincides with the proton cut point $2'$, which is shown in Fig.\ref{f3} for $A^p(-\omega,k)$ and in Eq.~(\ref{201}).
 So, we are forced  to conclude that or 1) $\omega_{sp}(k)$ at $k< k^p$ is placed under the both (1,1') and (2,2')  cuts in Fig.\,\ref{f3} for $A^p$ and we must take into account $2p2h$ states, or 2) accept that at $k< k^p$ the collective excitation  $\omega_{sp}(k)$ annihilates  with $ph$ pairs. Study is in progress.

\subsection{The branches $\omega_{snp}(k,\beta)$}
The branches $\omega_{snp}(k,\beta)$ are formed by the solutions of Eq.(\ref{8z}) just like $\omega_{sn}(k,\beta)$ and $\omega_{sp}(k,\beta)$. This equation coincides with  Eq.(\ref{8s}) if $\beta=0$. As it discussed above,  Eq.(\ref{8s}) has a branch of solutions $\omega_{s}(k)$, it is real at $k\leq k_t$ and complex at $k>k_t$. The branch $\omega_{snp}(k,\beta)$ is defined as continuation of the complex part of $\omega_{s}(k)$ to  $\beta\neq 0$.

In Fig.\,\ref{f6}$c$ and \ref{f7} we show $\omega_{snp}(k)$ for different $\beta>0$; $\omega_{snp}(k)$ starts at $k=k^{np}$ and is placed on the unphysical sheets only. At $\beta=0$ we have $k_t=k^{np}$ (Fig.\,\ref{f5}).

As it is shown in Fig.\,\ref{f5} the function $k^{np}(\beta)$ consist of two parts $k^{np}_{1}(\beta)$ and $k^{np}_{2}(\beta)$. The behavior of $\omega_{snp}(k)$ (shown in Fig.\,\ref{f8})  explains  the behavior of $k^{np}(\beta)$. At $\beta<0.24$ there is no solutions at small $k$, they exist at $k \geq k^{np}_1(\beta)$ only. But at $0.24\leq \beta \leq 0.37$ the branch $\omega_{snp}(k,\beta)$ consists of two parts. For example, at $\beta=0.3$ (Fig.\,\ref{f8})  the first part exists at $0\leq k \leq k^{np}_{2}(\beta)$ and the second one at $k > k^{np}_{1}(\beta)$. For $\beta > 0.37$ the branch $\omega_{snp}(k,\beta)$ exists at all $k$ (relevant to our approach). The imaginary part of $\omega_{snp}(k,\beta)$ corresponds to the decay  of $\omega_{snp}(k)$ due to mixture  with nucleon $ph$ pairs, but isospin of these pairs is not fixed unlike $\omega_{sn}(k)$ and $\omega_{sp}(k)$.

\subsection{Behavior of solutions at $|\beta| \to 1$}
As it was explained in Introduction, we wait a special behavior of solutions at $\beta>0$ and $p_{Fp}/p_0\to 0$ (and $\beta<0$ and $p_{Fn}/p_0 \to 0$).
As the density of the proton states go to zero we wait disappearance of  $\omega_{sp}(k)$ and   $\omega_{snp}(k)$. To take into account the decreasing of the density of the proton states we change the normalization of the effective interaction, Eq.(\ref{3r}).
The dispersion equation for the zero sound in ANM is presented in Eq.(\ref{8z}). In this subsection we compare solutions at $|\beta|\sim 1$ of the Eqs.(\ref{8}) and (\ref{8z}).

The presentation of results is given in Fig.\,\ref{f9}.  We fix $k=p_0$ and change the asymmetry parameter $\beta$.  The  numbers ${\it 1}$ and ${\it 2}$ denote the solutions of Eqs.(\ref{8}) and (\ref{8z}), correspondingly. We see that $\omega_{sp}(k)$ and  $\omega_{snp}(k)$ marked by ${\it 2}$ decrease with $\beta \to 1$ and only $\omega_{sn}(k)$ survives.

More detailed figures for the solutions ${\it 2}$ are presented in Figs.\ref{f10} for $\omega_{sp}(k)$. In the Figs.\ref{f10} ($left$) the real and imaginary parts of $\omega_{sp}(k)$ are shown as function of $\beta$. The dotted curve stands for $[{\rm Re}\,\omega_{sp}(k)-k^2/(2m)]/p_0$. The more convincing results are shown in Figs.\ref{f10} ($right$) where the dependence on $p_{Fp}/p_0$ of the same functions is demonstrated.  We conclude that  $[{\rm Re}\,\omega_{sp}(k)-k^2/(2m)]/p_0 \to 0$  at least as quick as  $p_{Fp}/p_0$ at $\beta\to1$, but  ${\rm Re}\,\omega_{sp}(k)/p_0$ do not. ${\rm Im}\,\omega_{sp}(k)/p_0$ go to zero as well.

Let's define $[{\rm Re}\,\omega_{sp}(k)-k^2/(2m)]\approx \alpha(k)\,p_{Fp}$, where $\alpha(k)$ is a complex function. Looking at Fig.\ref{f10}, we conclude that $\alpha(k)$ does not increase with $p_{Fp}$ decreasing.

We put in Eq.(\ref{10})  $\omega\approx k^2/(2m) + \alpha\,p_{Fp}$ and  take into account that $p_{Fp}/p_0\to 0$ then
\begin{equation}\label{14}
A^p=A^p(\omega,k)+A^p(-\omega,k)\approx
 -2\frac1{4\pi^2}\frac{m^3}{k^3}\left[p_{Fp}^2\,\left(\frac{\alpha^2-(k/m)^2}2\, \left[\ln\left(\frac{\alpha+k/m}{\alpha-k/m}\right) - 2\pi\,i\right] -\alpha\,k/m\right) \right.+
\end{equation}
$$
 \left.+\left(\frac{k^4/m^2 -b_p^2}2\,\ln\left(\frac{-k^2/m + b_p}{-k^2/m - b_p}\right) + b_p\,k^2/m\right)\right]=
$$
$$
= -2\frac1{4\pi^2}\frac{m^3}{k^3}p_{Fp}^2\, \left[\frac{\alpha^2-(k/m)^2}2\, \left[\ln\left(\frac{\alpha+k/m}{\alpha-k/m}\right)  - 2\pi\,i\right] - \alpha\,k/m  + \frac{2\,p_{Fp}\,k}{3\,m^2} +...\right].
$$
Here  a value $2\pi\,i$ appears due to the branch $\omega_{sp}(k)$ is treated on the unphysical sheet of $A^p(\omega,k)$.
This approximate results demonstrate that the term $C_{0p}\,F' A^p$ of the dispersion equation (\ref{8z}) turns to zero at $p_{Fp}/p_0\to 0$:  $C_{0p}\,F' A^p\approx p_{Fp}/p_0\phi(k,\alpha)$, where $\phi(k,\alpha)$ is a finite at  $p_{Fp}/p_0\to0$. Thus Eq.(\ref{8z}) becomes the dispersion equation for neutrons $ph$ pairs. The contribution of the proton $ph$ pairs disappears.

Let's compare disappearance of the proton solutions in the dispersion equations   Eqs.(\ref{8}) and (\ref{8z}), Fig.\ref{f9}. The branch of solutions  $\omega_{sp}(k)$ is placed on the unphysical sheet of function $A^p(\omega,k)$. When $p_{Fp}/p_0\to 0$  the proton cut
 $(1,1')$, Eq.(\ref{201}), go to  one point $\omega\to k^2/(2m)$. The branch $\omega_{sp}(k)$ of  Eq.(\ref{8}) (numbers ${\it 1}$ in Fig.\ref{f9}) remains on the unphysical sheet, and its physical sense is unclear when $p_{Fp}/p_0\to0$. On the other hand, if we consider $\omega_{sp}(k)$ as solutions of (\ref{8z}), then ${\rm Re}\,\omega_{sp}(k)\to k^2/2m$ (Figs.\ref{f10}). In this case it looks, in a limit,  as the knock out proton moves as a free one with the energy equal to its kinetic energy. This is not really so if $p_{Fp}/p_0\neq 0$   since $\omega_{sp}(k)$ is placed on the unphysical sheet and    ${\rm Im}\,\omega_{sp}(k)$ is a very essential  that $\omega_{sp}(k)$ be a solution of (\ref{8z}) at $p_{Fp}/p_0\neq 0$.

Situation with $\omega_{snp}(k,\beta)$ is close to that for $\omega_{sp}(k,\beta)$.

\subsection{Behavior of solutions at $|\beta| \to 0$}
We demonstrate the behavior of branches  $\omega_{sn}(k)$, $\omega_{sp}(k)$ and $\omega_{snp}(k)$ at small $|\beta|$. We deal with $\beta > 0$ mainly.

At $\beta=0$ in SNM we have two branches  $\omega_{s}(k)$ and  $\omega_{s1}(k)$ (Fig.\,\ref{f2}). Fig.\,\ref{f4} is made for  small $\beta=0.01$. In this figure we see  redistribution of $\omega_{s}(k)$ and $\omega_{s1}(k)$ by $\omega_{sn}(k)$,  $\omega_{sp}(k)$ and $\omega_{snp}(k)$.\\
 At $k\leq k_t(\beta)$ (on the left of '1', Fig.\ref{f4}$a$)  $\omega_{s}(k)$ is real and passes into the real part of  $\omega_{sn}(k)$ (compare figures $a$ and $b$).\\
At $k>k_t(\beta)$ (on the right of '1', Fig.\ref{f4}$a$) $\omega_{s}(k)$ is complex and transits to $\omega_{snp}(k)$ (compare figures $a$ and $d$). In figures $a$, $b$, $c$ the branch $\omega_{s1}(k)$  looks like splitting on $\omega_{sn}(k)$ and $\omega_{sp}(k)$ which are placed on the different unphysical sheets.

At first, we consider the dependence of the real solutions of Eq.(\ref{8z})  on $\beta$. They are shown at $k<k_t(\beta)$ in Figs.\,\ref{f2}, \ref{f4}($a,b$), \ref{f5}. The real solutions  are the lines started at $k=0$ and $\omega=0$ and finished at $k=k_t(\beta)$ and $\omega(k_t,\beta)$ (Fig.\,\ref{f5}). As explained above we attribute them at $\beta=0$ to  $\omega_{s}(k)$: $\omega(k,\beta)=\omega_s(k,\beta=0)$ and at $\beta>0$ to  $\omega_{sn}(k,\beta)$: $\omega(k,\beta)=\omega_{sn}(k,\beta>0)$. We see that there is the continuous transition with change of  $\beta$ between the different types of branches at $k\leq k_t(\beta)$ .
The branch $\omega_{sn}(k,\beta)$  is the continuation of $\omega_{s}(k,\beta=0)$ to $\beta> 0$ when $\omega_{s}(k,\beta=0)$ is real, Figs.\ref{f4},\ref{f5}.

Now let consider  the branch $\omega_{snp}(k)$,  which by definition is the continuation of the complex part of $\omega_{s}(k)$ to $\beta\neq0$.
In  Fig.\,\ref{f8} at small $\beta$ the branches $\omega_{snp}(k)$ start at $k_{1}^{np}(\beta)$ (see Fig.\,\ref{f5}) and $k_{1}^{np}(\beta)\to k_t(\beta=0)$ when $\beta\to 0$.  In Fig.\,\ref{f8} we see that at $\beta\to 0$ branches $\omega_{snp}(k)$ go to $\omega_{s}(k)$ (solid curve) (see the next subsection).

In Fig.\,\ref{f11} the branches $\omega_{sn}(k)$, $\omega_{sp}(k)$ and $\omega_{s1}(k)$ are shown on the complex $\omega$-plane. This presentation is a very sensitive to the distinction of branches.  The branch $\omega_{s1}(k)$ shown by the solid curve in Fig.\,\ref{f11} corresponds to the dashed curve in Fig.\,\ref{f4}$a$.
In Fig.\,\ref{f11} we see that the branches $\omega_{sn}(k)$ and $\omega_{sp}(k)$ calculated for $\beta$=0.01 envelope  $\omega_{s1}(k)$ (number '1'). This is more expressive for smaller $\beta$ (but difficult to draw since they are too close). For the larger $\beta$: $\beta=0.1$ (number '2'), the branches do not  feel $\omega_{s1}(k)$ at all.  This  explains the distinct behavior of these branches at small $\beta$, Figs.\,\ref{f6},\ref{f7}. So we conclude that $\omega_{s1}(k)$ is the limit of $\omega_{sn}(k)$ and $\omega_{sp}(k)$ at $\beta\to 0$ (from different unphysical sheets).
This is true at $k> k_c$ when $\omega_{s1}(k)$ exists since we do not obtain  the solutions of $\omega_{sn}(k)$ or  $\omega_{sp}(k)$ type at $\beta=0$  and $k<k_c$.  Another presentation of this limit is discussed in Fig.\ref{f13}.

Our results demonstrate how the solutions at $\beta=0$: $\omega_{s}(k)$, $\omega_{s1}(k)$ are continued to $\beta\neq 0$. We deal with $\omega_{s}(k)$ as consisting of two parts: real (at $k\leq k_t$) and complex (at $k\geq k_t$). At $\beta>0$ the real part turns into  the real part of $\omega_{sn}(k,\beta)$  but the complex part turns into $\omega_{snp}(k,(\beta))$. The branch  $\omega_{s1}(k)$ exist at $k>k_c$ and it is continued at $\beta\neq 0$ into two unphysical sheets, splitting into $\omega_{sn}(k,\beta)$ and $\omega_{sp}(k,\beta)$. One unphysical sheet belongs to $A^n(\omega,k)$, the second one to $A^p(\omega,k)$, Eq.(\ref{10}).

\subsection{Presentation  of the solutions at $-1\leq \beta\leq 1$}
In this section we demonstrate the behavior of solutions $\omega_{sn}(k,\beta)$, $\omega_{sp}(k,\beta)$, $\omega_{snp}(k,\beta)$ and $\omega_{spn}(k,\beta)$ when asymmetry parameter changes in the total interval  $-1\leq \beta \leq 1$ (Figs.~\ref{f12},{\ref{f14}). Our investigations indicate that the behavior differs for the  branches of the different types and it depends on value of $k$.

At the beginning we present the  branches $\omega_{sn}(k,\beta), \omega_{sp}(k,\beta)$  Figs.\,\ref{f12}. We explained above that there are the special values of the wave vector $k_t(\beta)$, $k^p(\beta)$, $k_c$, $k^{np}(\beta)$, Fig.\,\ref{f5}, which determine the change of branches with $\beta$.
 In Fig.\ref{f12} we demonstrate $\beta$-dependence of the real and imaginary parts of branches $\omega_{sn}(k)$, $\omega_{sp}(k)$  at some values $k_1=0.05p_0$, $k_2=0.4p_0$, $k_3= 0.6p_0$, these $k_1, k_2, k_3$ are placed in the three different intervals: $k_1 < k_t(\beta)$, $k_t(\beta) <k_2<k_c$ and $k_3> k_c$.

In the upper part of Fig.\ref{f12} the real parts of solutions, ${\rm Re}\,\omega_{si}(k,\beta)$, are shown. At $k=k_1$ (solid curves)  and $\beta>0$ the branch  $\omega_{sn}(k_1)$ is real. Changing $\beta$ from $\beta=1$ to negative values we go to the real $\omega_{sp}(k_1)$ at $\beta<0$.  At $\beta=0$ the branch goes through the point $\omega_s(k_1)$.
 At $k=k_1$  and $\beta>0$ the branch  $\omega_{sp}(k_1)$ is complex, it does not exist (at least, it is not found) at $k<k^p(\beta)$ and can not be continued to negative $\beta$. The same we can say at $\beta<0$ about $\omega_{sn}(k_1)$ which can not be continue to positive $\beta$.

At $k=k_2$ (the dotted curves in Fig.\ref{f12}) and $\beta>0$ the branch $\omega_{sn}(k_2)$ is complex (see Figs.\ref{f6},\ref{f7}). As explained above we can not continue it to $\omega_{sp}(k_2,\beta<0)$ changing $\beta$ from the positive to negative values. There is the discontinuity in dependence of the solutions on $\beta$  at  $\beta=0$ and  $k=k_2$. 

 At $k=k_3$ (the dashed curves in Fig.\ref{f12})  and $\beta>0$  the branch $\omega_{sn}(k)$ is a complex one but we have a complex solution at $\beta=0$ ($\omega_{s1}(k_3)$). We can move $\beta$ to negative values  and continue  $\omega_{sn}(k_3,\beta>0)$   to  $\omega_{sp}(k_3,\beta<0)$ through the point $\omega_{s1}(k_3)$.

In Fig.\,\ref{f12} (lower part)  the imaginary parts of branches $\omega_{sn}(k,\beta)$, $\omega_{sp}(k,\beta)$ for the same $k_1, k_2, k_3$ are shown. At $k= k_1$ the branch $\omega_{sn}(k,\beta>0)$ and $\omega_{sp}(k,\beta<0)$ are real by the definition of $k_1$.
At $\beta>0$  ${\rm Im}\,\omega_{sp}(k_1,\beta)$ is very small (see Sect.(4.2)) and tends to zero at $k\to k^p$ (but does not reach it since it exists at the unphysical sheet).
At $k=k_2$ at $|\beta| \to 0$ the imaginary parts of branches go to zero, but do not reach the point $\beta=0$ since there are not solutions of the  dispersion equation at $\beta=0$ for these $k$, Fig.\,\ref{f2}.
At $k=k_3$  ${\rm Im}\,\omega_{sn}(k,\beta>0)$ transit to ${\rm Im}\,\omega_{sp}(k,\beta<0)$ when we move $\beta$ to negative values.

In Figs.\ref{f12} for every $k$ we see how the  imaginary parts of $\omega_{sp}(k,\beta)$ ($\omega_{sn}(k,\beta)$) go to zero when $\beta\to 1$ ($\beta\to -1$) and the real parts go to $k^2/(2m)$ (see discussion in Sect.4.4).

Another presentation of Fig.\ref{f12} is given in Fig.\ref{f13}.  We demonstrate how $\omega_{sn}(k,\beta > 0)$ turns into $\omega_{sp}(k,\beta < 0)$ for different $k$. In the every point above the  axis $\beta=0$ we have solution $\omega_{sn}(k,\beta)$. Now let decrease $\beta$ to negative values. The dashed vertical lines with the arrows demonstrate the direct way down at $k<k_t$ and $k>k_c$. But there is not the way for $k_t < k< k_c$ since, as it was mentioned, there is not solutions of $\omega_{s\tau}(k)$, $\tau=n,p$ type in this interval of real axis. It looks as a boundary in this interval. To reach points below the boundary we must go around the point $k=k_c$ (shown by the dotted curve). It is possible to go around the point $k=k_t$, as well.

In Fig.\,\ref{f14}  we present the dependence of $\omega_{snp}(k,\beta)$ and $\omega_{spn}(k,\beta)$ on $\beta$ at three different points  $k_1=0.3\,p_0, k_2=0.355\,p_0, k_3=0.4\,p_0$.  Looking at the curve $k^{np}(\beta)$ in Fig.\,\ref{f5} we expect the different dependence of $\omega_{snp}(k,\beta)$ on $\beta$ at these $k_{1,2,3}$ since  $k_1<k_t(\beta)$, $k_t<k_2<k_1^{np}(max)$ and  $k_3>k_1^{np}(max)$.
In Figs.\,\ref{f14} we see that at $k=k_1$  the branch $\omega_{snp}(k)$ starts at $\beta> \beta_1$ where $k_1^{np}(\beta_1)=k_1$. This correspond to Fig.\,\ref{f5}: there is not solutions on the left and below the curve $k_1^{np}(\beta)$.   At $k=k_2$ there are two parts of solutions: at a small $\beta\simeq0.02$ and  $\beta>\beta_2$, where $k_1^{np}(\beta_2)=k_2$.
In Fig.\,\ref{f14} the small part of solutions is shown at small $\beta$ and the main part at $\beta>\beta_2$.
 At $k=k_3$ we have solutions at all $\beta$ (see Figs.\,\ref{f5},\ref{f8}). So when we go from positive $\beta$ to negative ones,  the branch $\omega_{snp}(k)$  transits to $\omega_{spn}(k)$ crossing the point $\omega_{s}(k=0.4p_0)$ at $\beta=0$. See Figs.\,\ref{f14} for  the real and imaginary parts (solid curves).

As a result,  at $k=k_3$ we have solutions at all $\beta$. At $k=k_1$ there are not solutions for small $\beta$ and there is a broken dependence of $\omega_{snp}(k,\beta)$ at $k=k_2$. In all cases we see that $\omega_{snp}(k,\beta)$ and  $\omega_{spn}(k,\beta)$ decrease
 at $|\beta|\to 1$.

\section{Summary}

We consider nuclear matter with asymmetry parameter  $-1\leq \beta\leq 1$.
 We obtain the complex branches of the solutions of the zero sound dispersion equation (\ref{8z}). The imaginary part of them describes the decay of zero-sound excitations  due to an admixture  of the free particle-hole pairs (of a corresponding type) into the collective excitations.

The aim of this paper is to obtain the branches of zero sound solutions and to tie them for different $\beta$ and $k$. This determines the appearance and the type  of solutions.

The symmetric, asymmetric and neutron matter are examined.  In SNM we obtain two branches of solutions $\omega_{s}(k,\beta=0)$ and $\omega_{s1}(k,\beta=0)$ (Fig.\,\ref{f2}). In ANM  three branches are found: $\omega_{sn}(k,\beta)$, $\omega_{sp}(k,\beta)$ and $\omega_{snp}(k,\beta)$ for every $\beta>0$ (Figs.\,\ref{f6},\ref{f7}) and one branch is in the neutron matter $\omega_{sn}(k,\beta=1)$ (the dotted curve in Fig.\,\ref{f6}$a$). To obtain solutions for $\beta<0$ we should to change $n \leftrightarrow p$ in the notation of  the branches for $\beta>0$.

The changes of solutions with $\beta$ depend on the value of wave vectors. There are four values of $k$ which determine the different regions. For small $\beta$ we have $k^p(\beta) < k_t(\beta) \leq k_1^{np}(\beta) < k_c$ (Fig.\ref{f5}).

At $k<k_t(\beta)$ the branches $\omega_{sn}(k,\beta)$ (at $\beta>0$) and  $\omega_{sp}(k,\beta)$ (at $\beta<0$) are real and can be calculated for all $\beta$ (Fig.\,\ref{f12}) and it is easy to pass from one $\beta$ to another. We attribute the real solutions to the branches of the different type: for $\beta=0$ to $\omega_{s}(k)$, for $\beta>0$ to $\omega_{sn}(k)$ and to $\omega_{sp}(k)$ for $\beta<0$.  They are the continuation of the real function $\omega_{s}(k,\beta=0)$ on other $\beta$.

 At $k > k_t(\beta)$ solutions become complex. An additional branch $\omega_{snp}(k)$ is obtained. It is the continuation of the complex  part of  $\omega_{s}(k)$ to $\beta\neq 0$.

We investigated the behavior of branches when $|\beta|\to0$ and $|\beta|\to 1$. It is shown that at $\beta\to 1$ the imaginary parts of $\omega_{sp}(k)$ and  $\omega_{snp}(k)$ go to zero and ${\rm Re}\,\omega_{sp}(k)$, ${\rm Re}\,\omega_{snp}(k)$ go to $k^2/(2m)$. Only $\omega_{sn}(k)$ survives as zero sound excitation. We obtain this result in our calculations if to  take into account the decreasing to zero of the proton  density of states at $\beta\to 1$ (Eq.\,(\ref{1c0})).

At  $\beta\to0$ and  $k > k_t$  the complex part of $\omega_{s}(k)$ is the limit of the set of branches $\omega_{snp}(k,\beta)$.
When  $\beta\to 0$ and  $k > k_c$ the branches  $\omega_{sp}(k)$ and  $\omega_{sn}(k)$ go to $\omega_{s1}(k)$. At $k_t < k < k_c$ there is not the limit of $\omega_{sn}(k)$ and $\omega_{sp}(k)$  for $\beta\to0$, Fig.\ref{f13}. The solutions $\omega_{sn}(k,\beta>0)$ and $\omega_{sp}(k,\beta<0)$ look similar to the difference phases of state.

The results presented above  can be applied to the investigation of the response functions of the nuclear matter and  nuclei \cite{2004}. Note, that our results are very sensitive to the density and the quasiparticle interactions (even taken in the simple form Eq.(\ref{17}))  \cite{8}. The inclusion of the isoscalar interaction $F$ would demand separate consideration of excitations in the isoscalar channel. The previous results demonstrate that the mutual influence of isoscalar and isovector channels is weak.

\newpage

\begin{figure}%1
\centering{\epsfig{figure=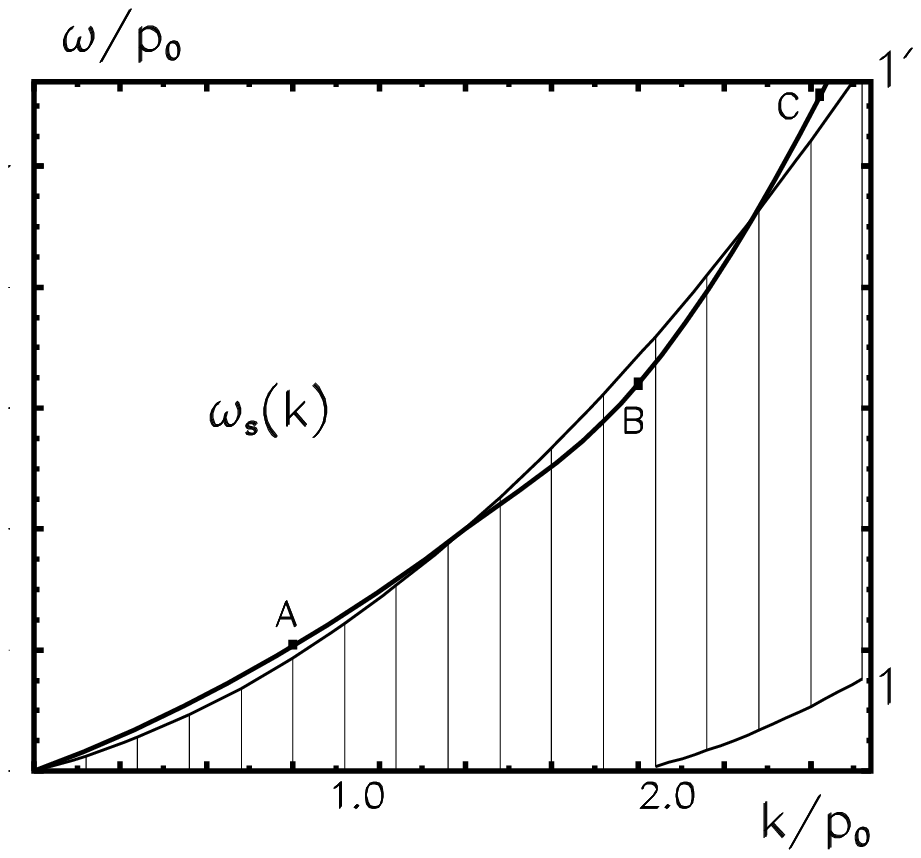,width=8cm}\epsfig{figure=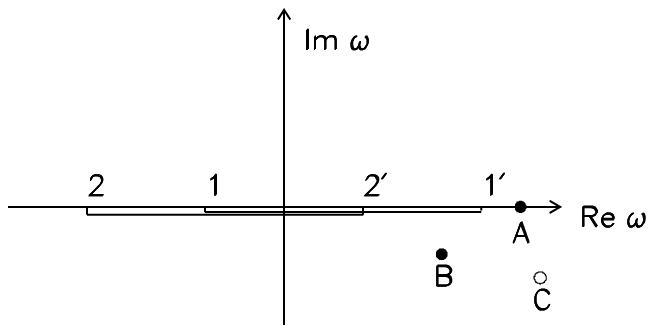,width=8cm}}
\caption{ A schematic figure. {\it left}:
The solid curve presents $\omega_{s}(k)$. The dashed area is occupied by the free $ph$ pairs $\omega_{ph}$ corresponding to the cut (1,1').  {\it right}:  The cuts of the  function A in (\ref{13},\ref{10}). The cut (1,1') is the cut of $A(\omega,k)$. The cut (2,2') is the cut of $A(-\omega,k)$. The point $A,B,C$  mark the solutions having the different location in respect to the cuts.
}
\label{f1}
\end{figure}

\begin{figure} %2
\centering{\epsfig{figure=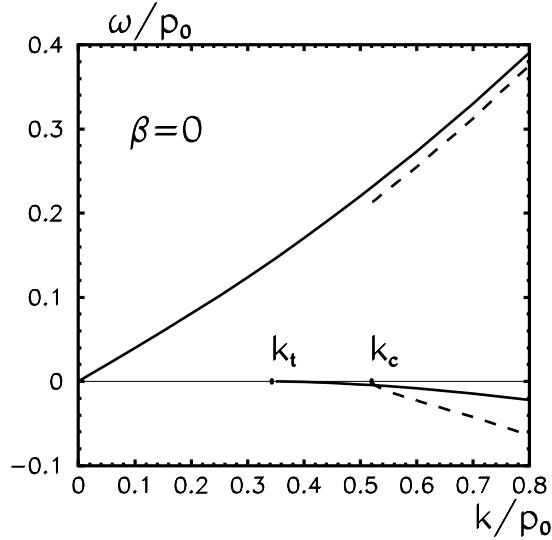,width=8cm}}
\caption{The branches of solutions in SNM. The solid  curves is for $\omega_{s}(k)$, the dashed curves - $\omega_{s1}(k)$. At $\omega>0$ we place the real parts of branches ${\rm Re}\,\omega_{i}(k)$ and at $\omega<0$ there are the imaginary parts:  ${\rm Im}\,\omega_{i}(k)$. Here $i=s, s1$.
}
 \label{f2}
\end{figure}

\begin{figure}%3
\centering{\epsfig{figure=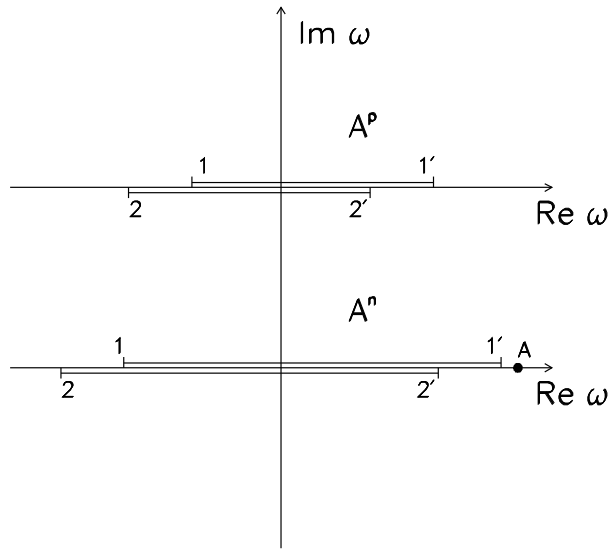,width=9cm}}
\caption{A schematic figure at $\beta>0$.  The cuts of the functions $A^\tau$ (\ref{10}) are presented on the complex $\omega$-plane. The cut (1,1') is the cut of $A^\tau(\omega,k)$. The cut (2,2') is the cut of $A^\tau(-\omega,k)$. In the upper (lower) part the  cuts of  $A^p$ ($A^n$) are demonstrated.
}
\label{f3}
\end{figure}

\begin{figure}%4
\centering{\epsfig{figure=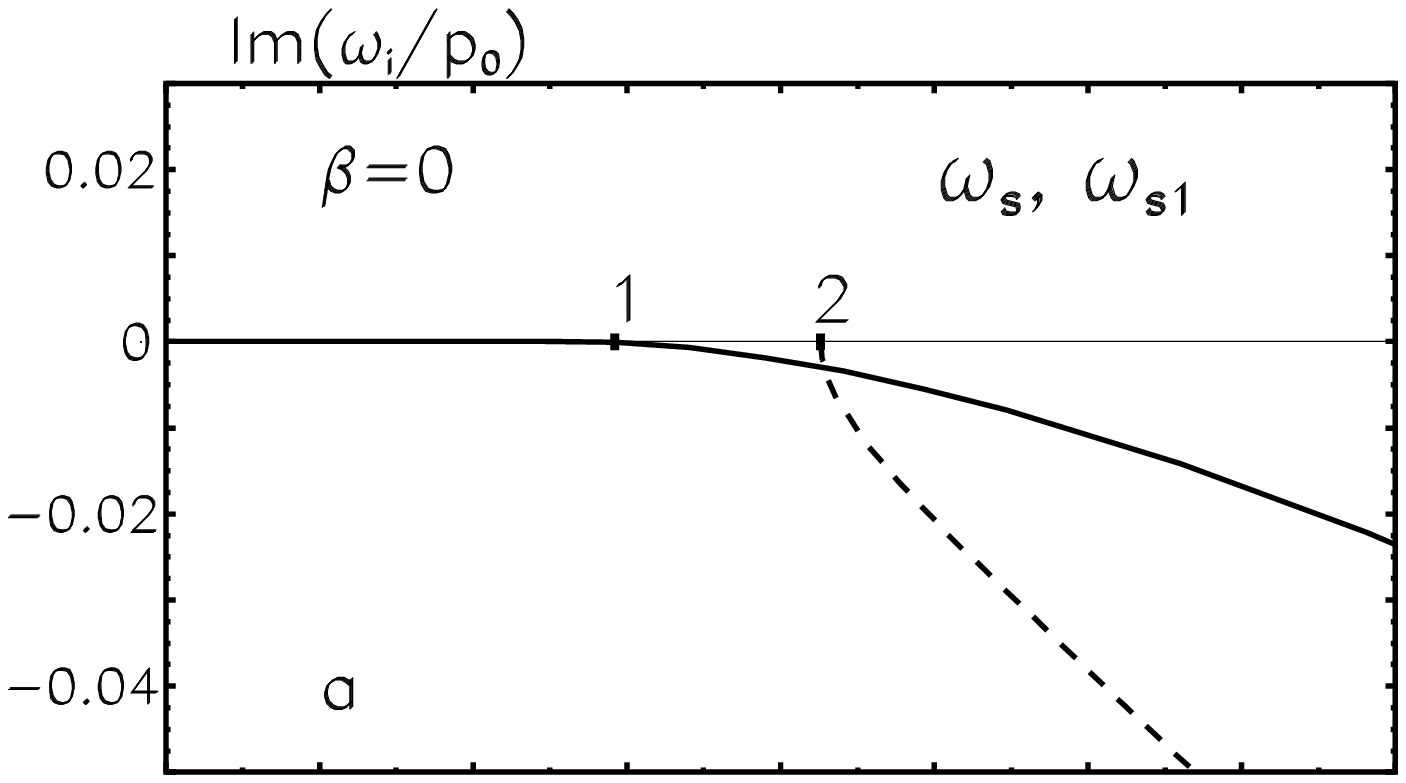,width=8.0cm}
\hspace{-1.4cm}\epsfig{figure=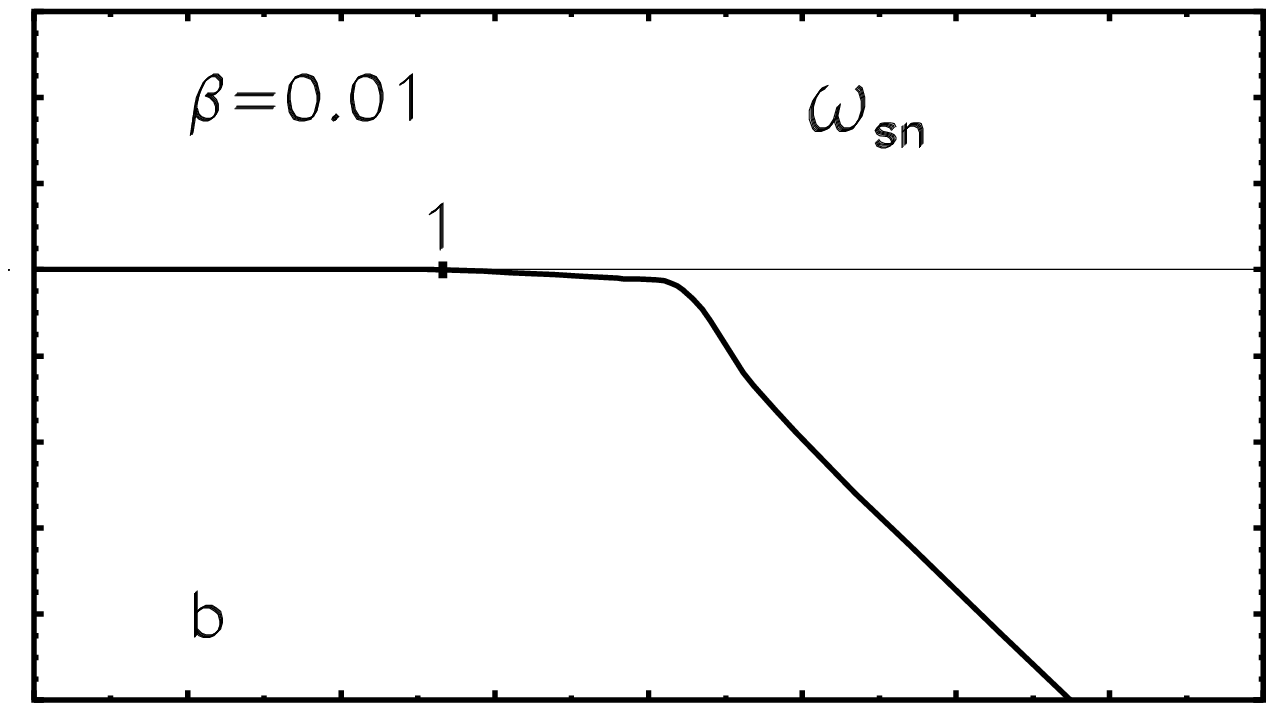,width=8.0cm}}

\vspace{-1.2cm}

\centering{\epsfig{figure=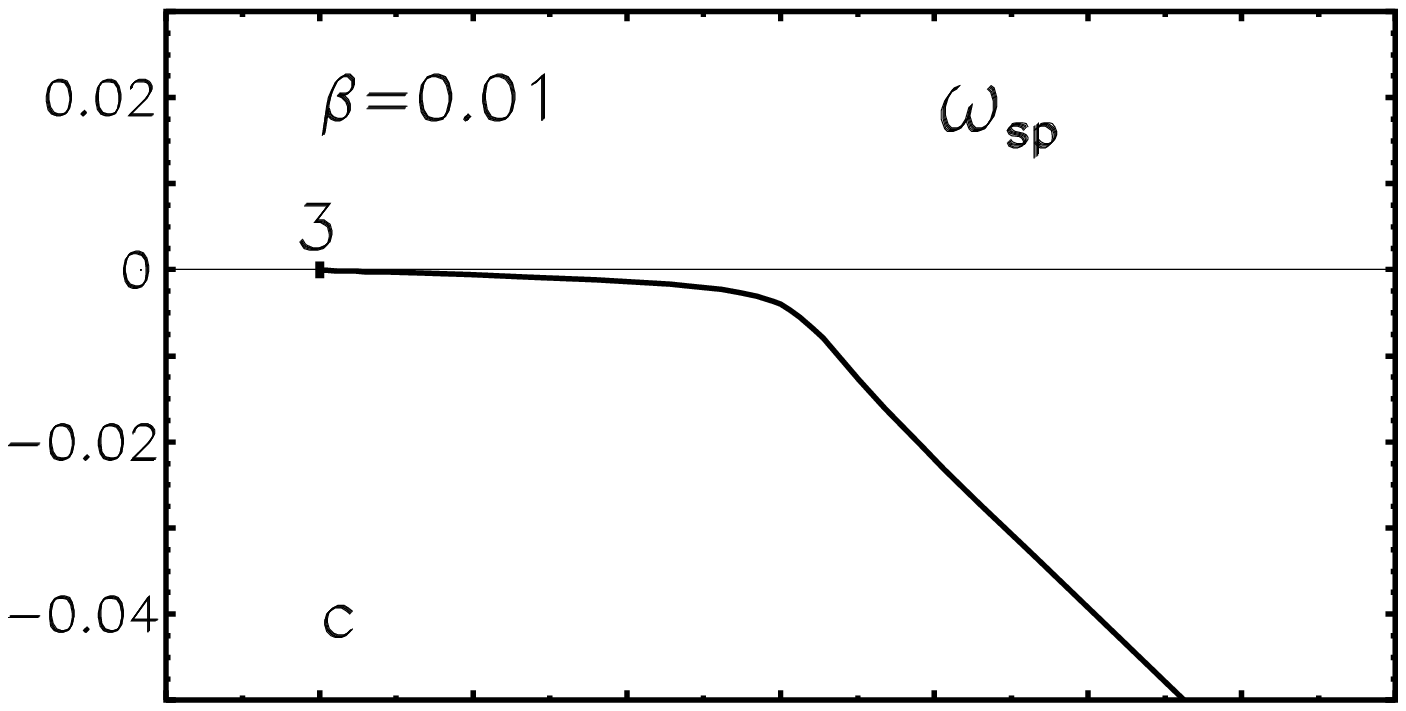,width=8.0cm}\hspace{-1.4cm}
\epsfig{figure=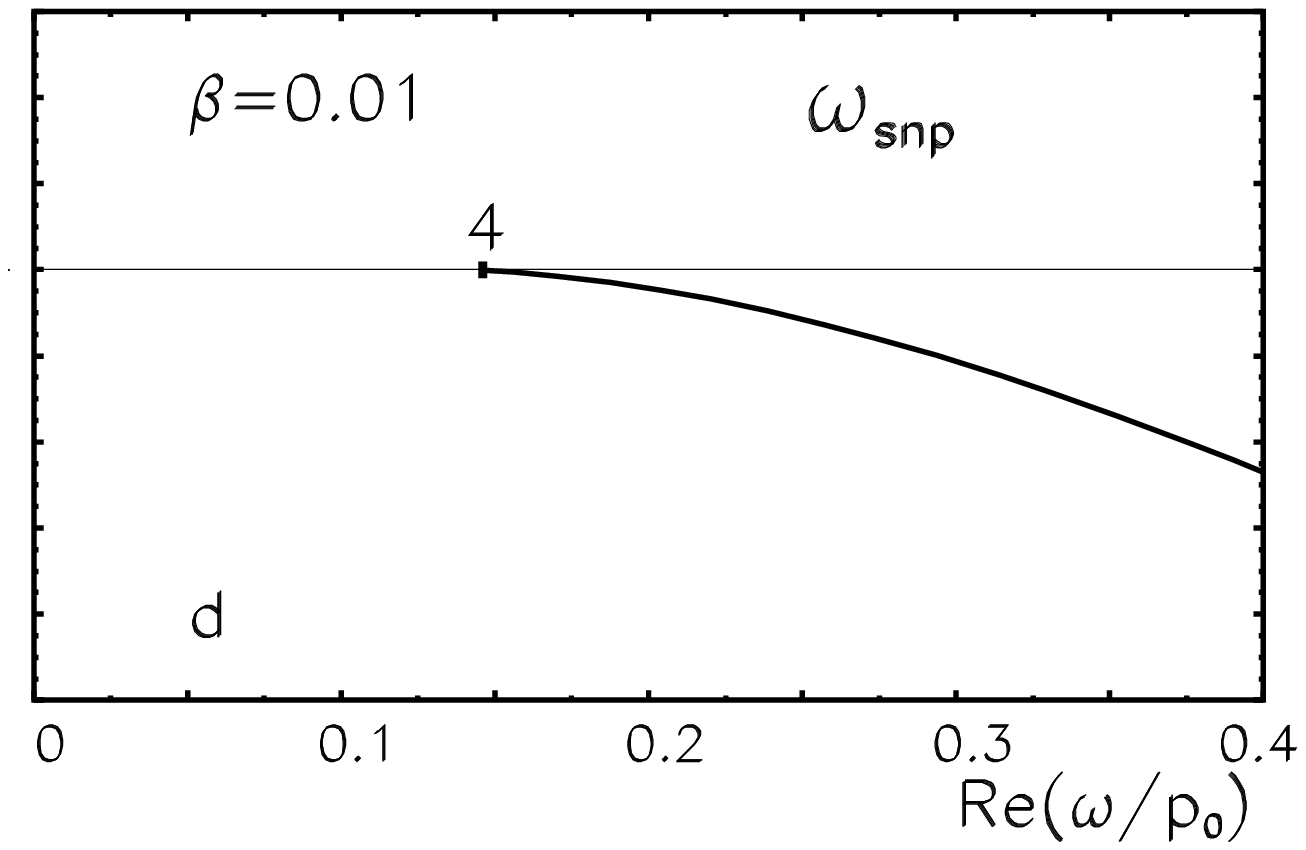,width=8.0cm}}
\caption{The complex $\omega$-plane. Comparison of zero-sound solutions in SNM and in ANM for small $\beta$. At the horizontal axis we show ${\rm Re}\,\omega_{si}(k)$, at the vertical axis  - ${\rm Im}\,\omega_{si}(k)$.
({\it a}): $\beta=0$, solid curve is $\omega_s(k)$, dashed curve is $\omega_{s1}(k)$; point '1' marks $\omega_s(k_t)$, '2' is  for $\omega_{s1}(k_c)$. In other figures $\beta=0.01$.
({\it b}):  the branch $\omega_{sn}(k)$, '1' stands for $\omega_{sn}(k_t)$.
({\it c}): the branch $\omega_{sp}(k)$,  point '3'  shows $\omega_{sp}(k^p)$.
({\it d}): the branch $\omega_{snp}(k)$, '4' - $\omega_{snp}(k^{np})$. According to Fig.\ref{f5} we have $k_c=0.52p_0$, $k_t(\beta=0)=0.34p_0$, $k_t(\beta=0.01)=0.32p_0$, $k^p(\beta=0.01)=0.15p_0$, $k^{np}(\beta=0.01)=0.35p_0$,
}
\label{f4}
\end{figure}

\begin{figure}%5
\centering{\epsfig{figure=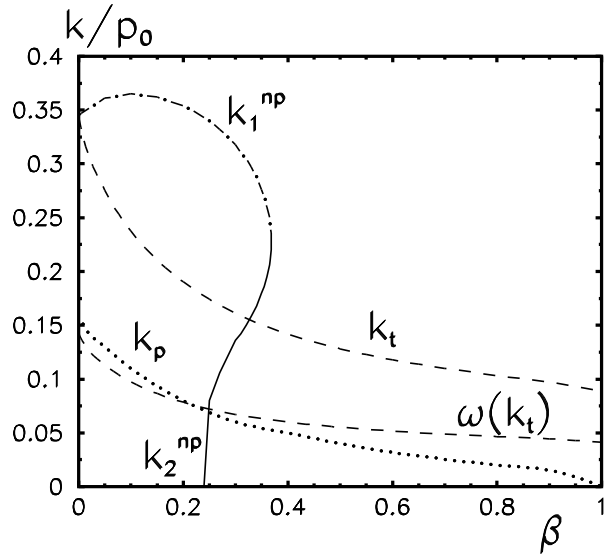,width=9cm}}
\caption{Dependence of a special wave vectors  $k_t(\beta)$, $k^p(\beta)$, $k^{np}(\beta)$ on $\beta$. The curve $k^{np}(\beta)$ consists from two parts: $k^{np}_{2}(\beta)$ (solid curve) and $k^{np}_{1}(\beta)$ (dash-dotted curve). The dashed curves  is for  $k_t(\beta)$; dotted curves - $k^p(\beta)$; $\omega(k_t)$ is the final point of the real solutions at different $\beta$. The curves can be specular reflected in a vertical axis $\beta=0$ to the negative $\beta$.}
\label{f5}
\end{figure}

\begin{figure} %6wtype
\centering{\epsfig{figure=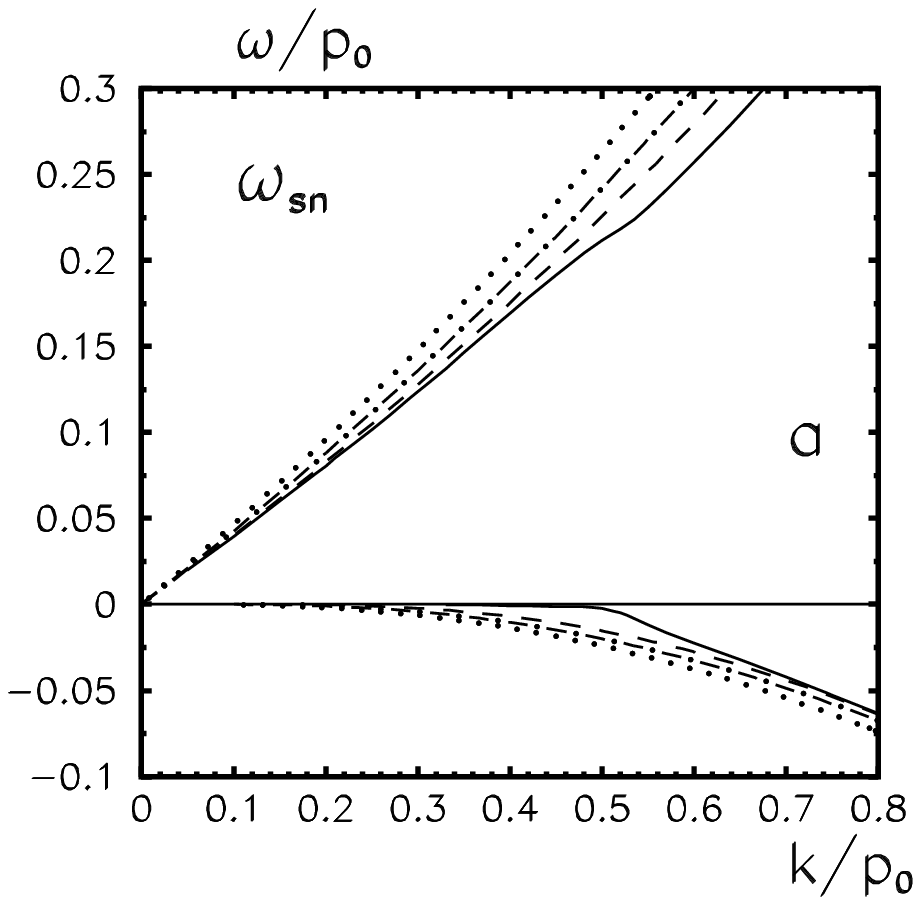,width=7cm} \epsfig{figure=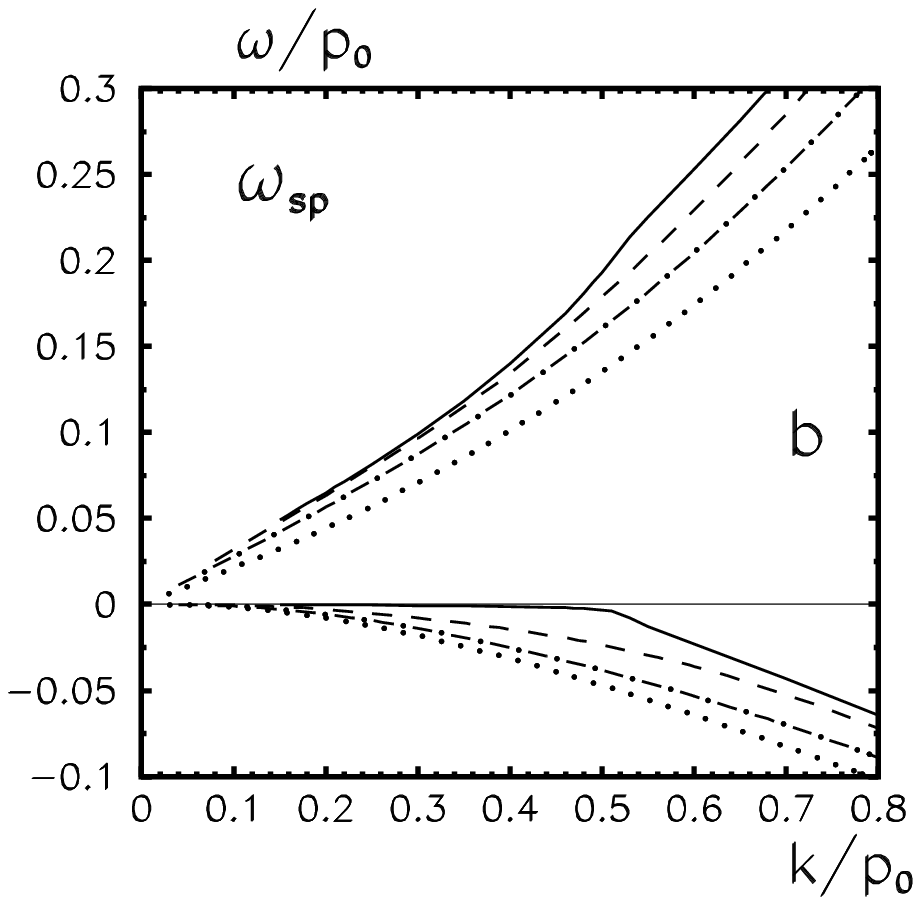,width=7cm}}
\centering{\epsfig{figure=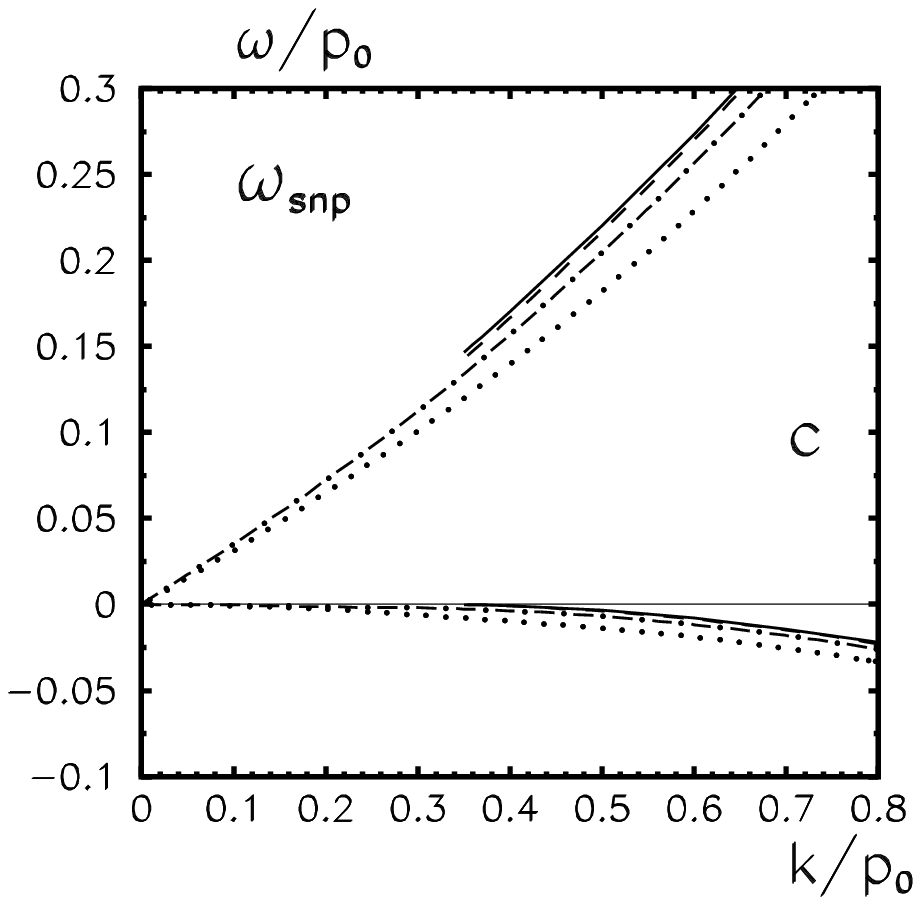,width=7cm}}
\caption{Dependence of the branches $\omega_{sn}(k)$ ($a$), $\omega_{sp}(k)$ ($b$) and $\omega_{snp}(k)$ ($c$)  on $\beta$. At $\beta$=0.01 -- solid curves; 0.2 -- dashed; 0.5 -- dot-dashed;  dotted curve: for $\beta=1.0$ in $a$ and $\beta=0.8$ in $b,c$. Other notations are the same as in Fig.~\ref{f2}. }
\label{f6}
\end{figure}

\begin{figure}%7wbeta
\centering{\epsfig{figure=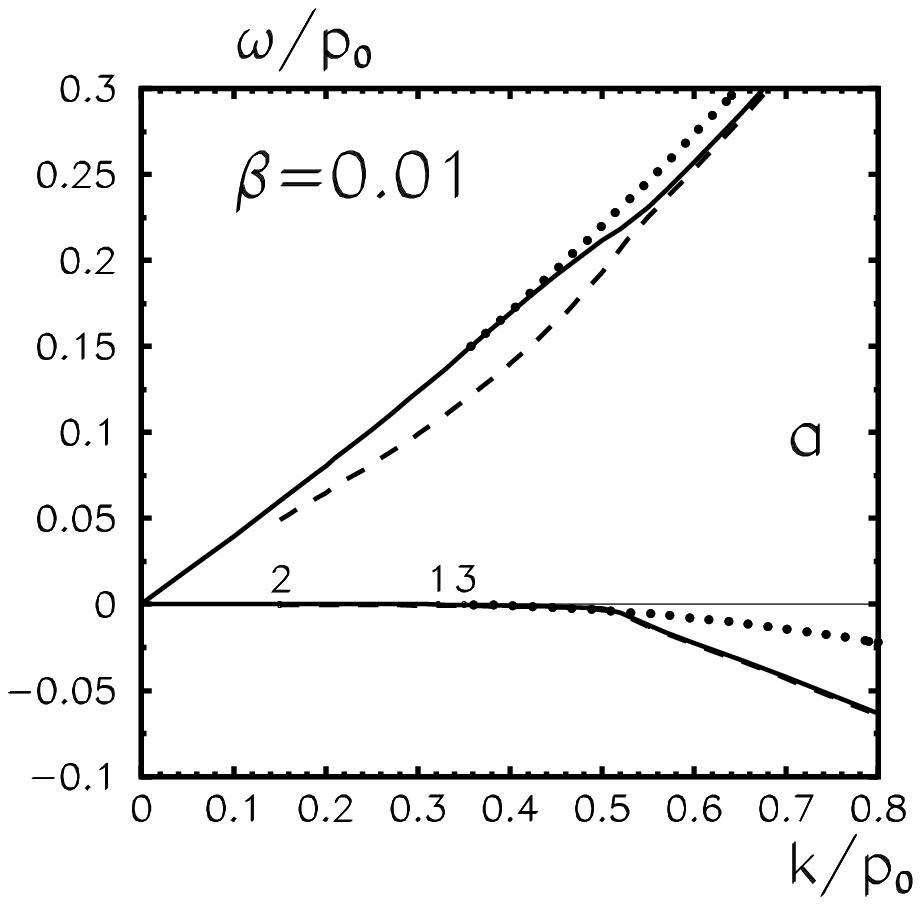,width=7cm} \epsfig{figure=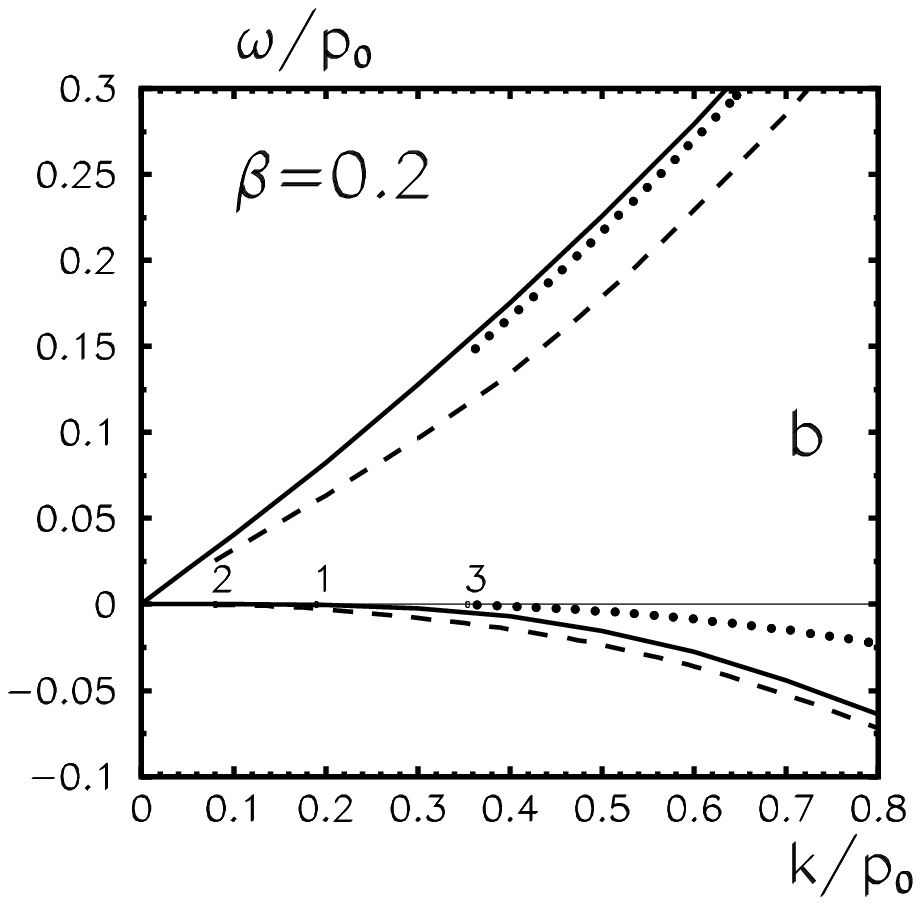,width=7cm}}
\centering{\epsfig{figure=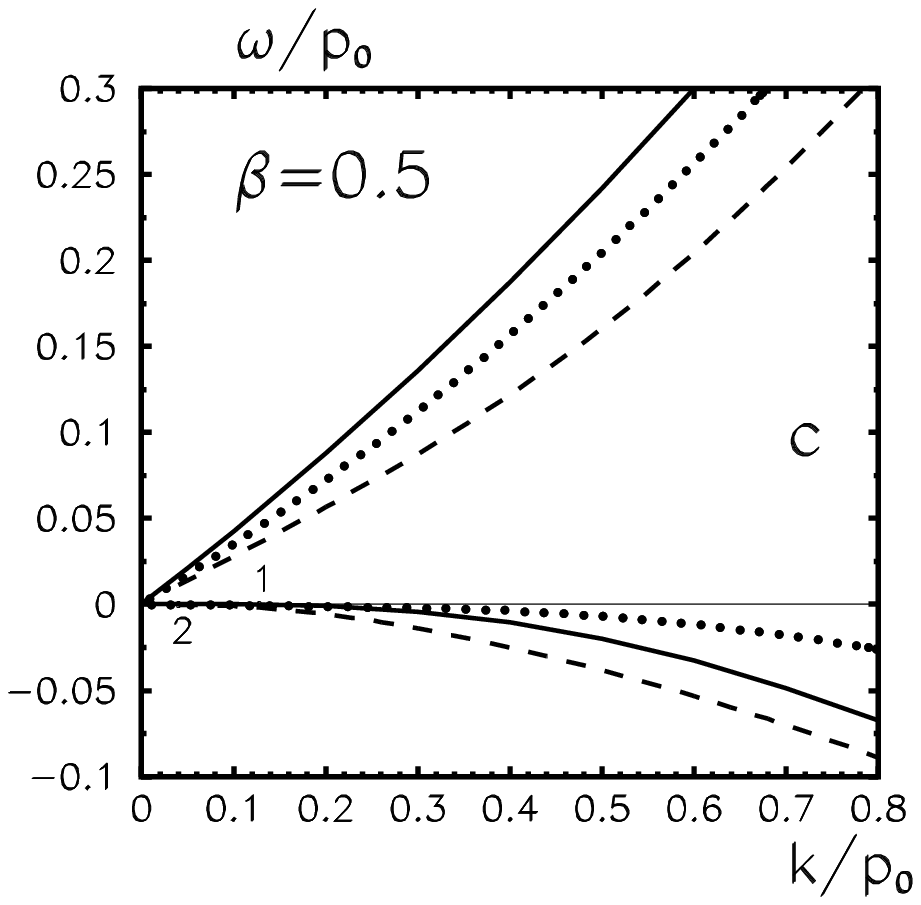,width=7cm} \epsfig{figure=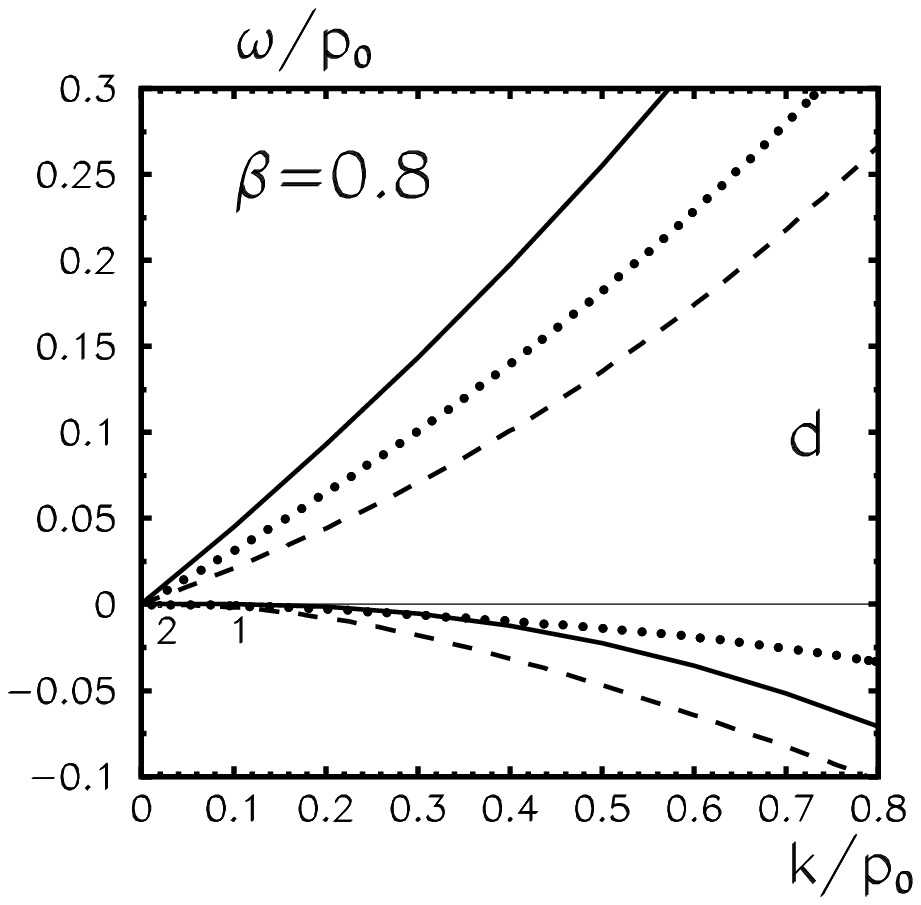,width=7cm}}
\caption{Comparison   $\omega_{sn}(k)$, $\omega_{sp}(k)$ and $\omega_{snp}(k)$   for different $\beta$: $\beta$=0.01 ($a$), 0.2($b$), 0.5($c$),  0.8($d$). The branches $\omega_{sn}(k)$ are shown by the solid curves; dashed curves are for $\omega_{sp}(k)$; dotted curves - $\omega_{snp}(k)$. The number '1' means wave vector $k_t(\beta)/p_0$, the number '2' - $k^p(\beta)/p_0$, '3' - $k^{np}(\beta)/p_0$. Other notations are the same as in Fig.~\ref{f2}.}
\label{f7}
\end{figure}

\begin{figure} %8
\centering{\epsfig{figure=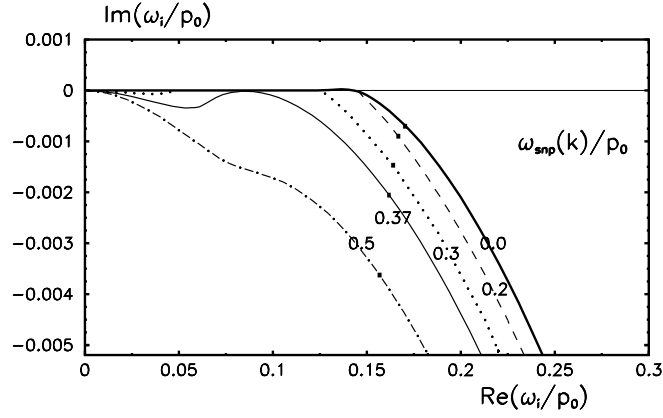,width=9cm}}
\caption{The complex $\omega$-plane. Presentation of  $\omega_{snp}(k,\beta)$ for different $\beta$. The values of $\beta$ are shown on the curves. The curve for $\beta=0$ is $\omega_s(k)$.
The point where $\omega_{snp}(k,\beta=0.37)$ touches the horizontal axis, takes place at $k=0.228p_0$. The fat dots stands at $k=0.4p_0$: $\omega_{snp}(k=0.4p_0,\beta)$.
}
 \label{f8}
\end{figure}

\begin{figure} %9
\centering{\epsfig{figure=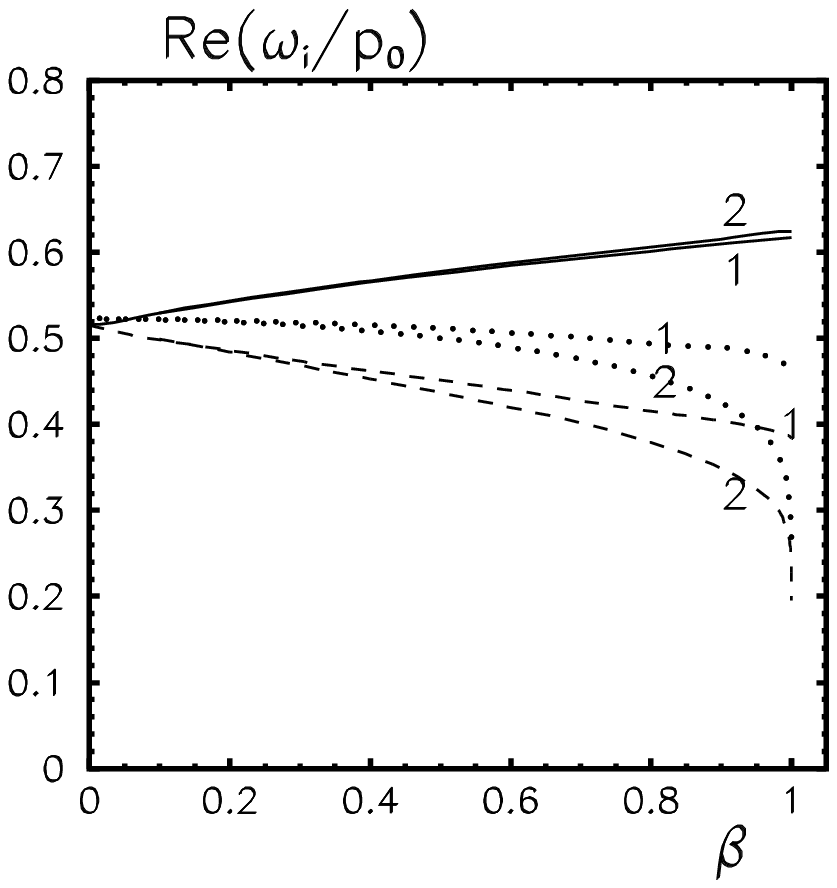,width=8cm}\epsfig{figure=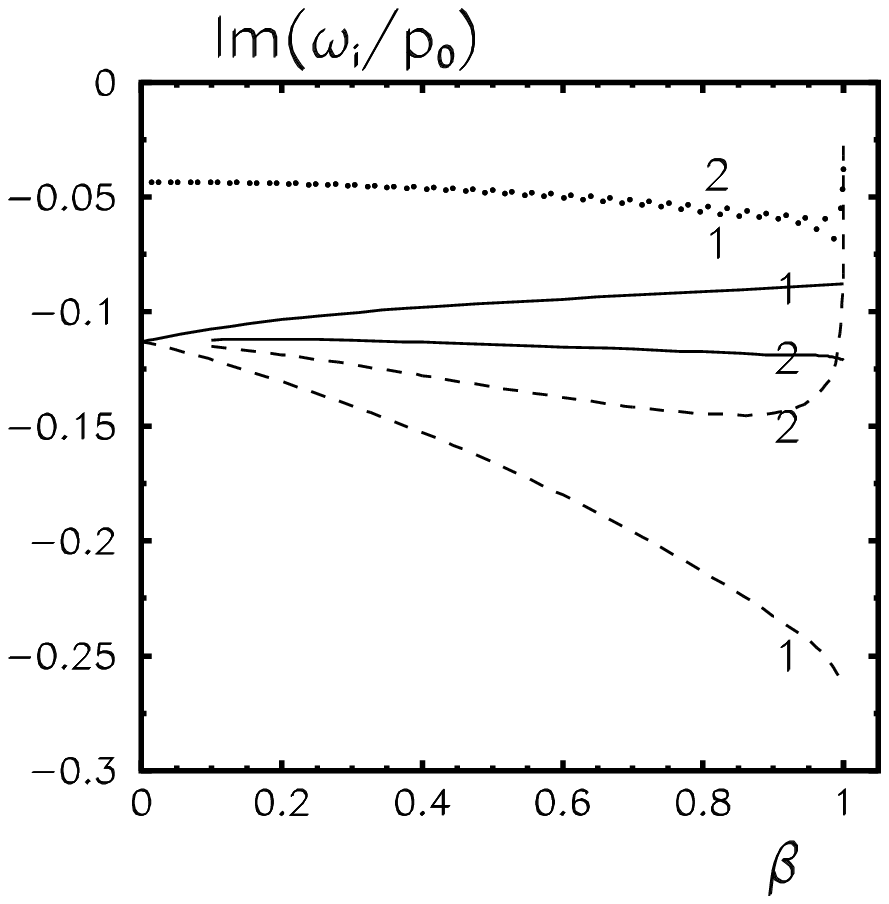,width=8cm}}
\caption{Behavior of $\omega_{sp}(k,\beta)$, $\omega_{sn}(k,\beta)$ and  $\omega_{snp}(k,\beta)$  at $\beta \to 1$. The wave vector $k$ is fixed: $k=p_0$. Solid curves denote $\omega_{sn}(k=p_0,\beta)$, dashed  curves -- $\omega_{sp}(k=p_0,\beta)$ and dotted curves are for $\omega_{snp}(k=p_0,\beta)$. The number '1' marks solutions of Eq.\,(\ref{8}) and '2' - solutions of Eq.\,(\ref{8z}).
 }
 \label{f9}
\end{figure}

\begin{figure}%10
\centering{\epsfig{figure=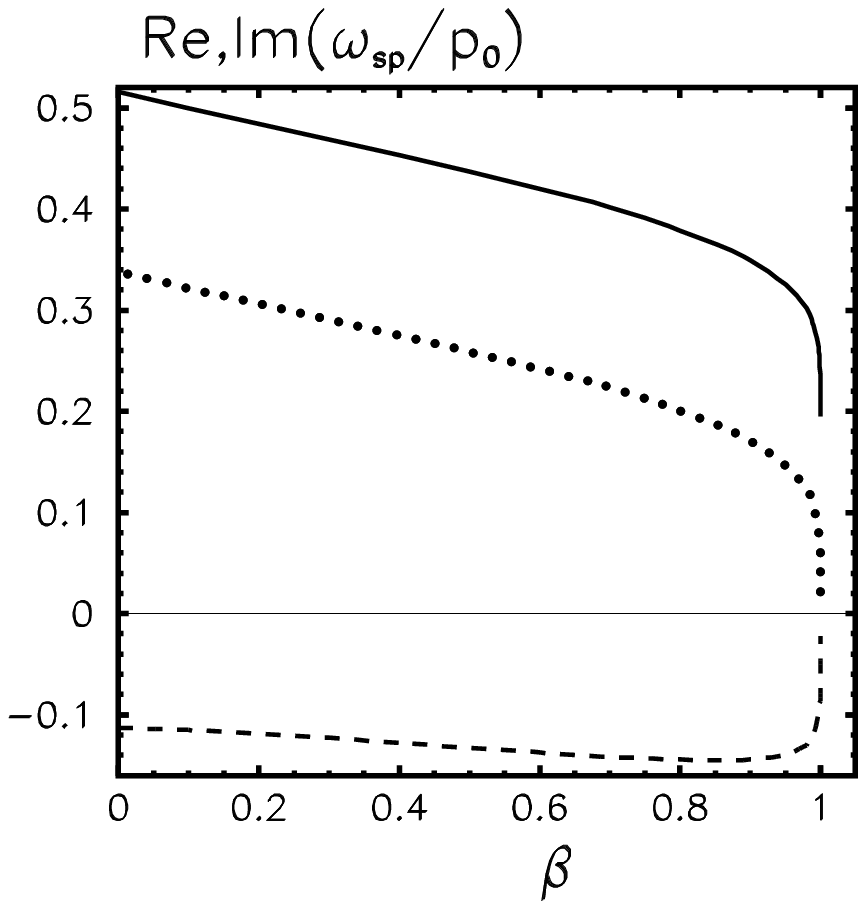,width=5.5cm}\epsfig{figure=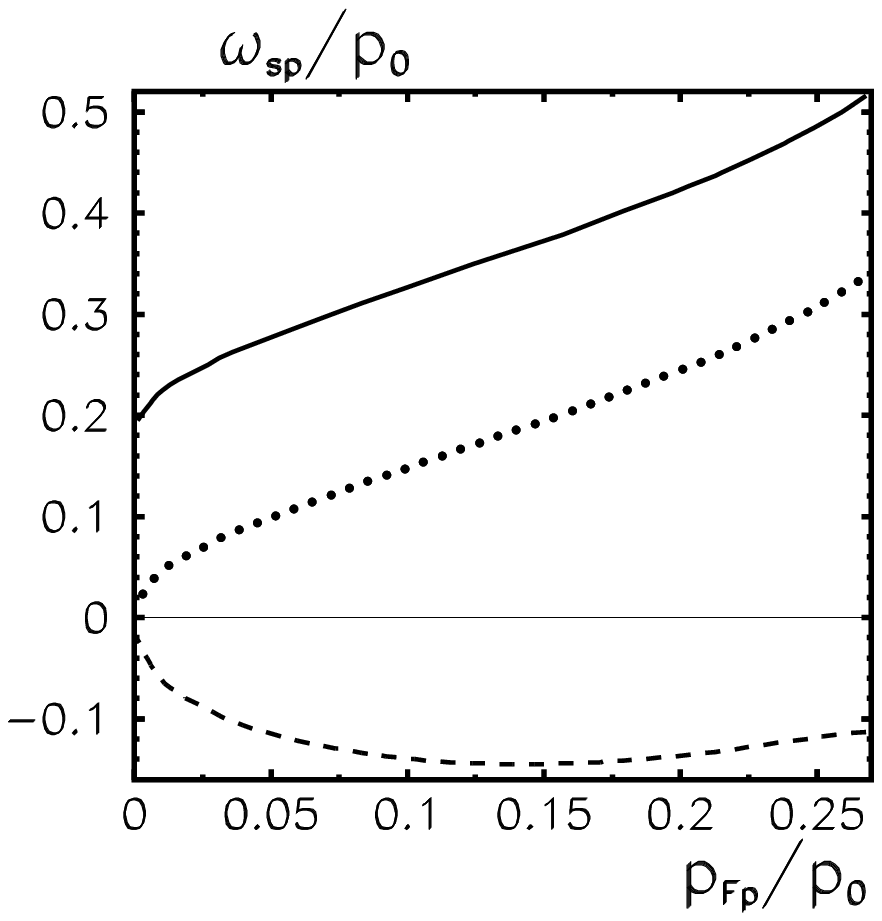,width=5.5cm}}
\caption{The same results as in Fig.\ref{f9}. $\omega_{sp}(k,\beta)$ at $\beta\to1$ and $k=p_0$ are shown from two point of views; ($left$): dependence on $\beta$; ($right$): dependence on $p_{Fp}$, corresponding to $\beta$; ${\rm Re}\,\omega_{sp}(k)$ is shown by  solid curve, the difference $[{\rm Re}\,\omega_{sp}(k)-k^2/(2m)]$ by the dotted curve, dashed curve is for ${\rm Im}\,\omega_{sp}(k)$.
}
\label{f10}
\end{figure}

\begin{figure} %11
\centering{\epsfig{figure=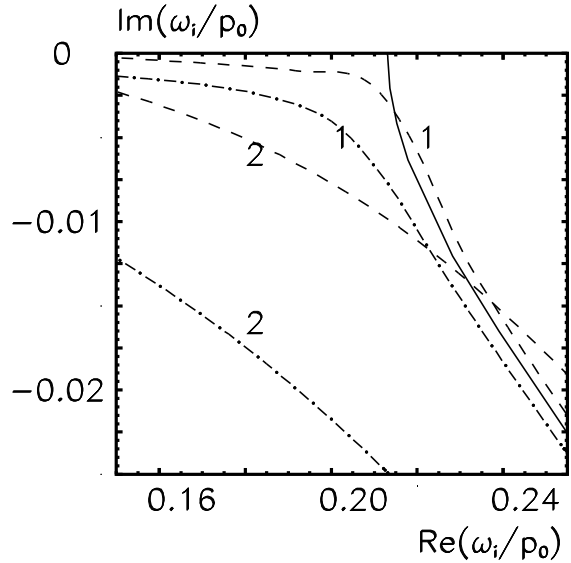,width=8cm}}
\caption{ The complex $\omega$-plane. Behavior of $\omega_{sp}(k,\beta)$, $\omega_{sn}(k,\beta)$ at $\beta \to 0$.
The branches $\omega_{sn}(k,\beta)$ (dashed) and  $\omega_{sp}(k,\beta)$ (dashed-dotted) for $\beta$=0.01 (noted by '1') and for $\beta$=0.1 (noted by '2')   in comparison with $\omega_{s1}(k)$ (solid) calculated for $\beta$=0.
 }
 \label{f11}
\end{figure}

\begin{figure}%12
\centering{\epsfig{figure=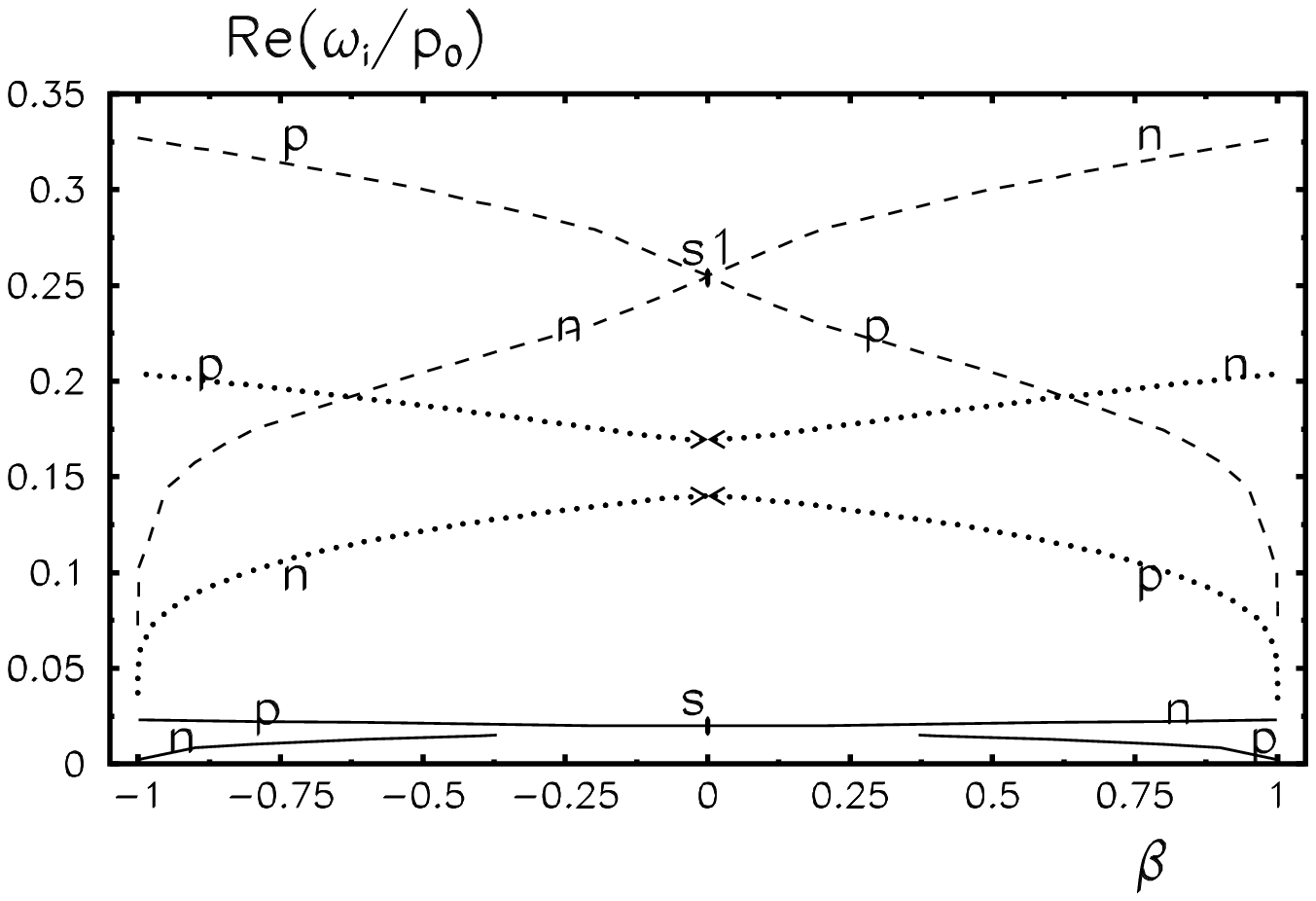,width=12cm}}
\centering{\epsfig{figure=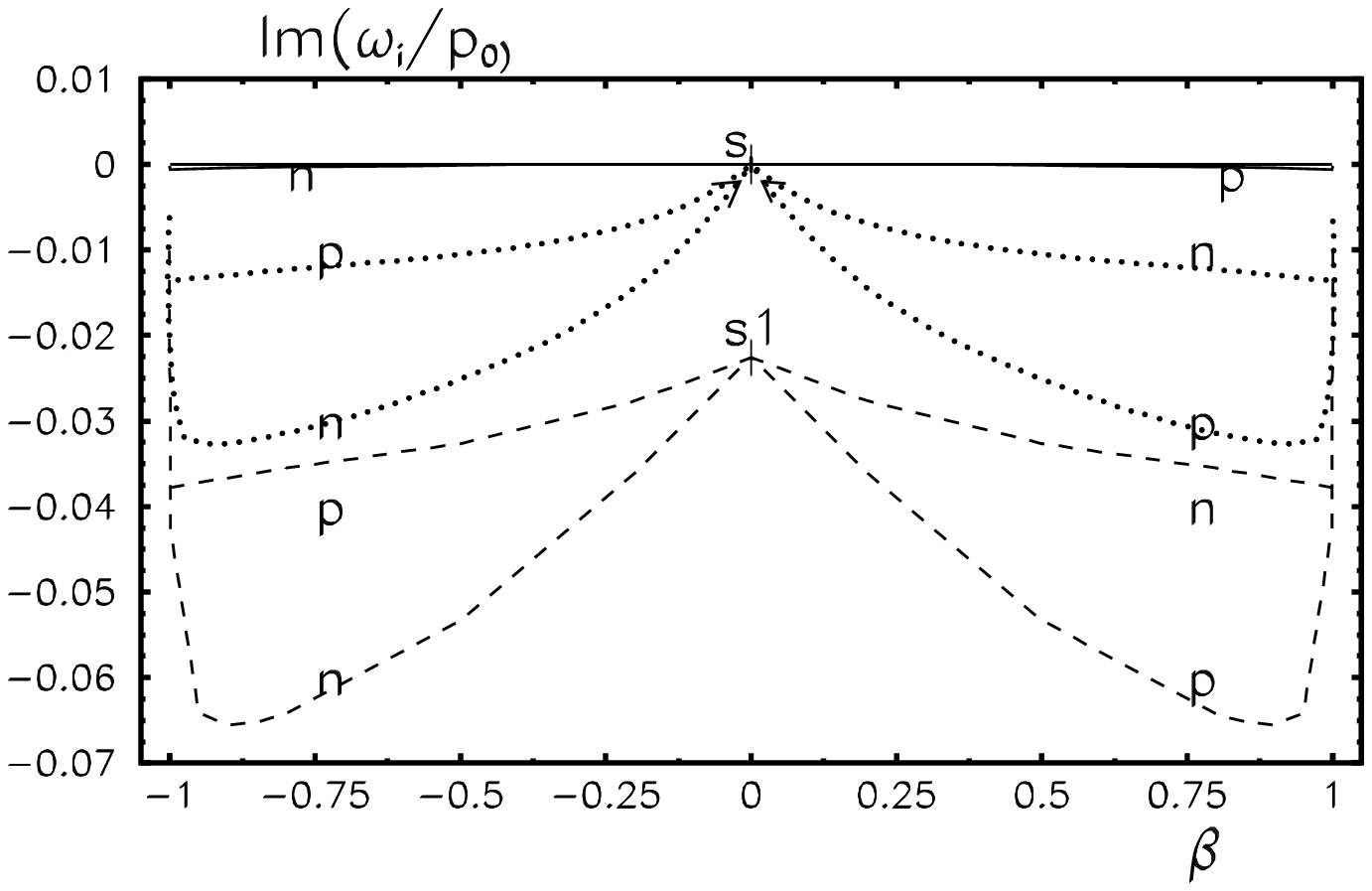,width=12cm}}
\caption{Dependence of the real and imaginary parts of $\omega_{sn}(k_i,\beta)$ and $\omega_{sp}(k_i,\beta)$ on $\beta$ at definite $k_i$, i=1,2,3.  Solid curves: $k_1=0.05p_0$, dotted lines: $k_2=0.4p_0$; dashed lines: $k_3=0.6p_0$;  `n` marks $\omega_{sn}(k_i,\beta)$, 'p'=$\omega_{sp}(k_i,\beta)$; 's'=$\omega_s(k_1)$, 's1'=$\omega_{s1}(k_3)$.}
\label{f12}
\end{figure}

\begin{figure} %13
\centering{\epsfig{figure=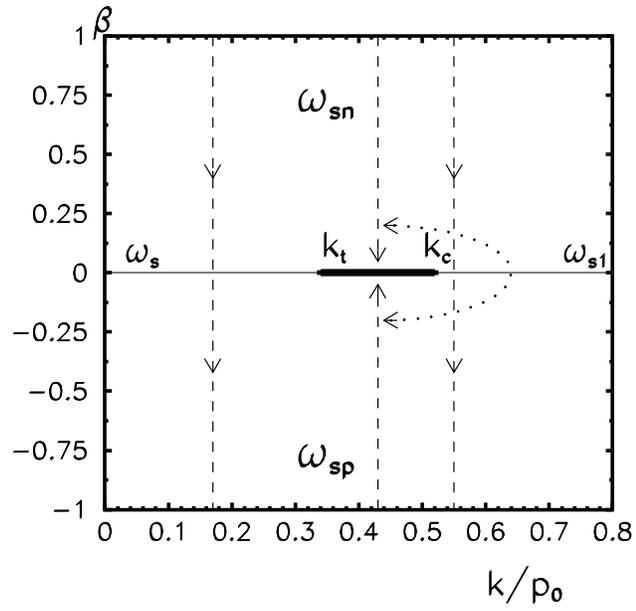,width=9cm}}
\caption{Another presentation of Fig.\ref{f11}. Branches  $\omega_{sn}(k)$ ($\omega_{sp}(k)$) are placed on the upper (lower) part of figure. The dashed lines with arrows demonstrate the continuous transition from $\omega_{sn}(k,\beta>0)$ into $\omega_{sp}(k,\beta<0)$ at $k\leq k_t$ and $k\geq k_c$. There is not the transition in the interval $k_t < k < k_c$. The dotted curve demonstrate the continuous way around this interval.
}
 \label{f13}
\end{figure}

\begin{figure}%14
\centering{\epsfig{figure=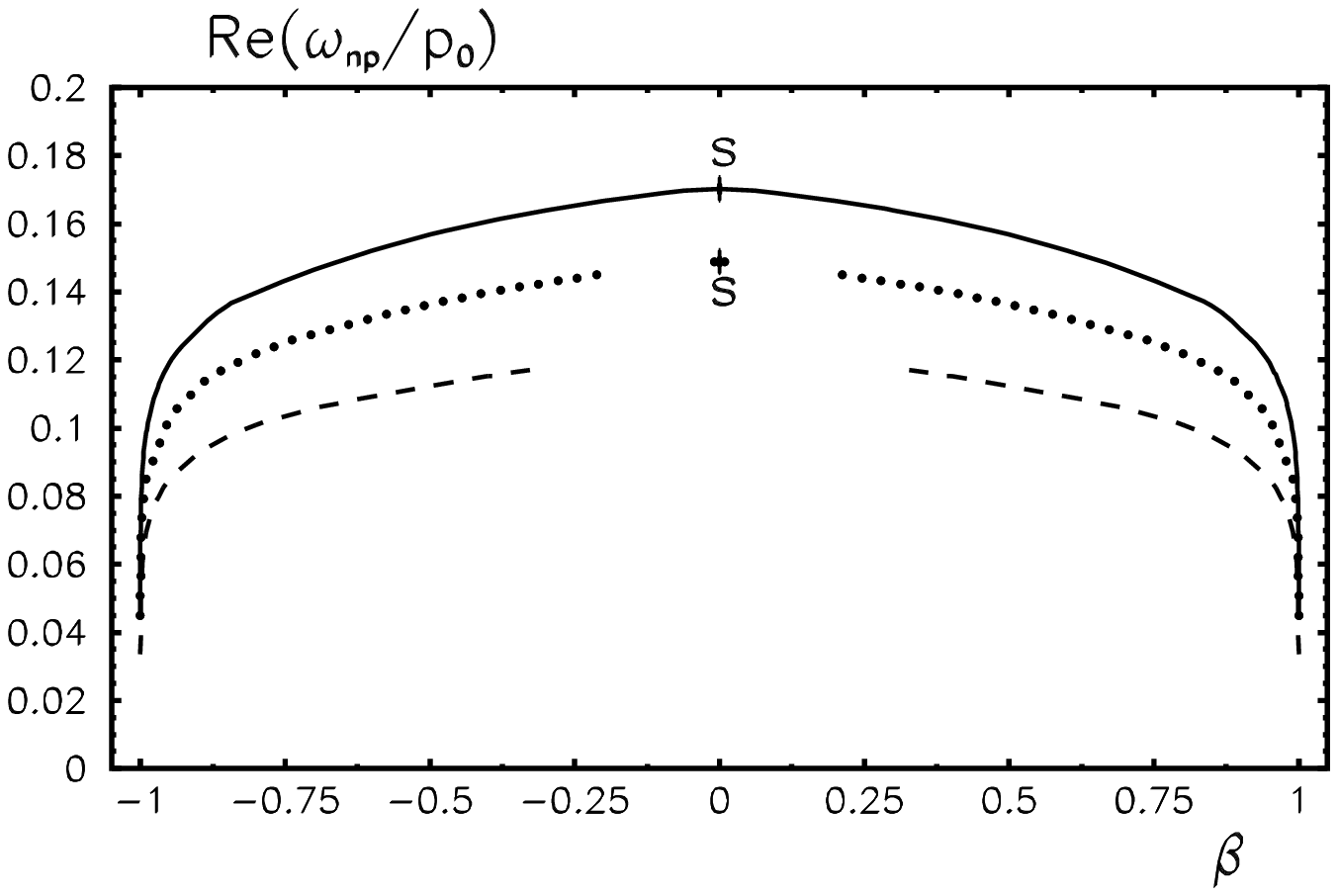,width=10cm}}
\centering{\epsfig{figure=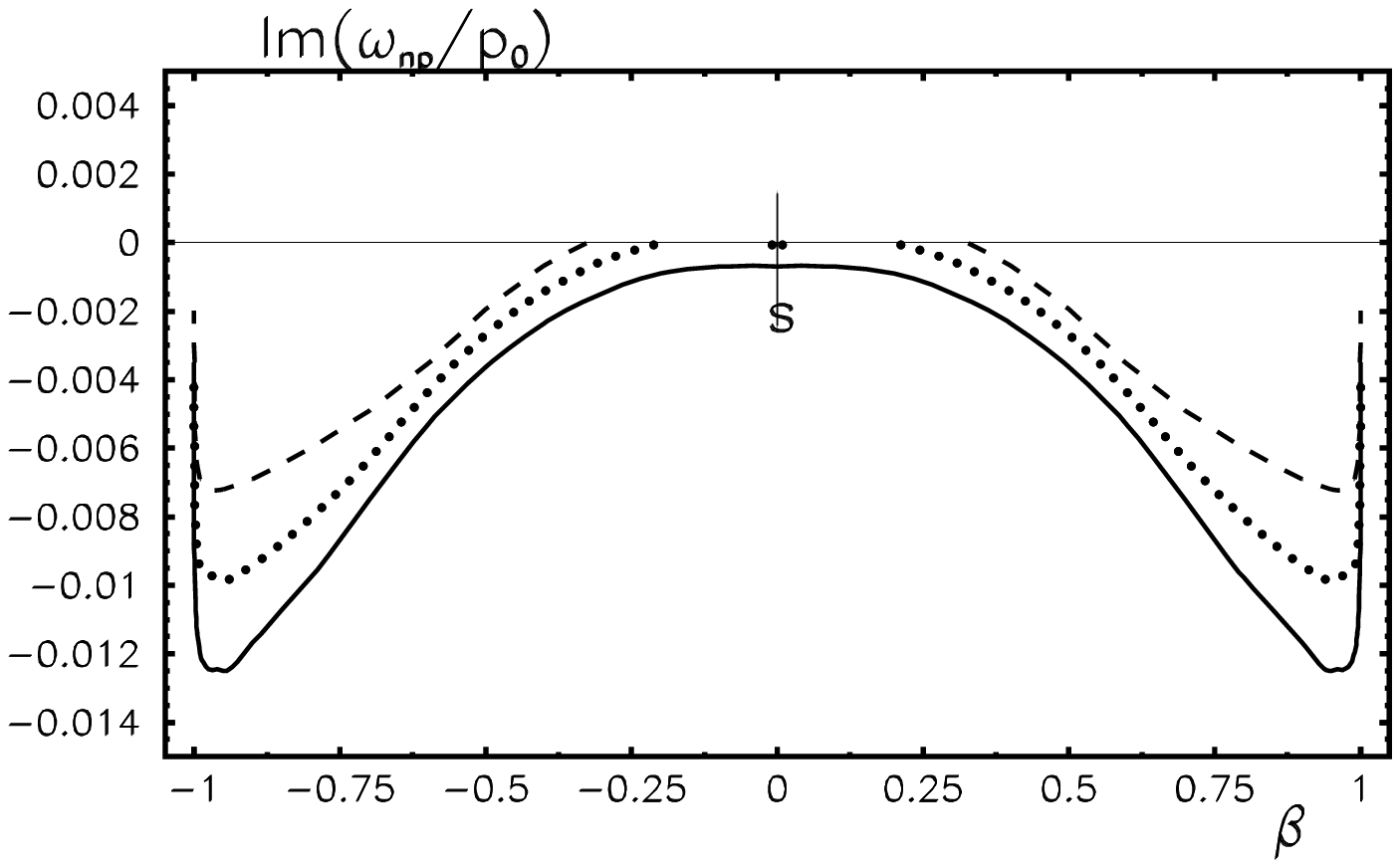,width=10cm}}
\caption{Dependence of the real and imaginary parts of $\omega_{snp}(k_i,\beta)$ and $\omega_{spn}(k_i,\beta)$ on $\beta$ at definite $k_i$, i=1,2,3.  Solid curves: $k_3=0.4p_0$, dotted curves: $k_2=0.355p_0$; dashed curves: $k_1=0.3p_0$; 's' marks $\omega_s(k_1)$ and $\omega_s(k_2)$.
At $\beta>0$ ($\beta<0$) the branch $\omega_{snp}(k_i,\beta)$ ($\omega_{spn}(k_i,\beta)$) is shown.
}
\label{f14}
\end{figure}

\end{document}